\documentclass[fleqn,usenatbib]{mnras}

\usepackage{newtxtext,newtxmath}

\usepackage[T1]{fontenc}
\usepackage{ae,aecompl}

\usepackage{graphicx}	
\usepackage{amsmath}	
\usepackage{amssymb}	
\usepackage{pdflscape}

\newcommand{\kms}{\mbox{km s$^{-1}$\space}}
\newcommand{\HCO}{\mbox{HCO$^{+}$}}

\title[Molecular Absorption in Hydra-A]{A molecular absorption line survey toward the AGN of Hydra-A}

\author[Tom Rose et al.]{Tom Rose$^{1}$\thanks{E-mail: thomas.d.rose@durham.ac.uk},
A. C. Edge$^{1}$, 
F. Combes$^{2}$,
S. Hamer$^{3}$,
B. R. McNamara$^{4}$,\newauthor
H. Russell$^{5}$,
M. Gaspari$^{6}$\thanks{\textit{Lyman Spitzer Jr.} Fellow.},
P. Salom\'e$^{2}$,
C. Sarazin$^{7}$,
G. R. Tremblay$^{8}$,\newauthor
S. A. Baum$^{9,10}$,
M. N. Bremer$^{11}$,
M. Donahue$^{12}$,
A. C. Fabian$^{13}$, \newauthor
G. Ferland$^{14}$,
N. Nesvadba$^{15}$,
C. O'Dea$^{9,16}$,
J. B. R. Oonk$^{17,18,19}$,
A. B. Peck$^{20}$
\\
\noindent $^{1}$Centre for Extragalactic Astronomy, Durham University, DH1 3LE, UK\\
$^{2}$LERMA, Observatoire de Paris, PSL Research Univ., College de France, CNRS, Sorbonne Univ., Paris, France\\
$^{3}$Department of Physics, University of Bath, North Rd, Bath, BA2 7AY \\
$^{4}$Department of Physics and Astronomy, University of Waterloo, Waterloo, ON N2L 3G1, Canada\\
$^{5}$Centre for Astronomy \& Particle Theory, University of Nottingham, Nottingham, NG7 2RD, UK\\
$^{6}$Department of Astrophysical Sciences, 4 Ivy Lane, Princeton University, Princeton, NJ 08544-1001, USA\\
$^{7}$Department of Astronomy, University of Virginia, 530 McCormick Road, Charlottesville, VA 22904-4325, USA\\
$^{8}$Center for Astrophysics $|$ Harvard \& Smithsonian, 60 Garden St., Cambridge, MA 02138, USA\\ 
$^{9}$Department of Physics \& Astronomy, University of Manitoba, Winnipeg, MB R3T 2N2, Canada \\
$^{10}$Chester F. Carlson Center for Imaging Science, Rochester Institute of Technology, 84 Lomb Memorial Dr., NY 14623, USA\\
$^{11}$HH Wills Physics Laboratory, Tyndall Avenue, Bristol, BS8 1TL, UK\\
$^{12}$Physics \& Astronomy Department, Michigan State University, East Lansing, MI 48824-2320, USA\\
$^{13}$Institute of Astronomy, Cambridge University, Madingly Rd., Cambridge, CB3 0HA, UK\\
$^{14}$Department of Physics and Astronomy, University of Kentucky, Lexington, Kentucky 40506-0055, USA\\
$^{15}$Universit\'e C\^ote d'Azur, Observatoire de la C\^ote d'Azur, CNRS, Laboratoire Lagrange, Bd de l'Observatoire, \\ CS 34229, 06304 Nice cedex 4, France\\
$^{16}$School of Physics and Astronomy, Rochester Institute of Technology, 85 Lomb Memorial Drive, USA\\
$^{17}$SURFsara, P.O. Box 94613, 1090 GP Amsterdam, The Netherlands\\
$^{18}$ASTRON, Netherlands Institute for Radio Astronomy, 7990AA Dwingeloo, The Netherlands\\
$^{19}$Leiden Observatory, Leiden University, Niels Borhweg 2, NL-2333 CA Leiden, The Netherlands\\
$^{20}$Gemini Observatory, Northern Operation Center, 67-0 N. A'Ohoku Place, Hilo, HI, USA 
}

\date{Accepted XXX. Received YYY; in original form ZZZ}

\pubyear{2020}

\begin{document}
\label{firstpage}
\pagerange{\pageref{firstpage}--\pageref{lastpage}}
\maketitle

\begin{abstract}
We present Atacama Large Millimeter/submillimeter Array observations of the brightest cluster galaxy Hydra-A, a nearby ($z=0.054$) giant elliptical galaxy with powerful and extended radio jets. The observations reveal CO(1-0), CO(2-1), $^{13}$CO(2-1), CN(2-1), SiO(5-4), HCO$^{+}$(1-0), HCO$^{+}$(2-1), HCN(1-0), HCN(2-1), HNC(1-0) and H$_{2}$CO(3-2) absorption lines against the galaxy's bright and compact active galactic nucleus. These absorption features are due to at least 12 individual molecular clouds which lie close to the centre of the galaxy and have velocities of approximately $-50$ to $+10$ \kms relative to its recession velocity, where positive values correspond to inward motion. The absorption profiles are evidence of a clumpy interstellar medium within brightest cluster galaxies composed of clouds with similar column densities, velocity dispersions and excitation temperatures to those found at radii of several kpc in the Milky Way. We also show potential variation in a $\sim 10$ \kms wide section of the absorption profile over a two year timescale, most likely caused by relativistic motions in the hot spots of the continuum source which change the background illumination of the absorbing clouds. 
\end{abstract}



\section{Introduction}

\begin{figure*}
	\includegraphics[width=0.99\textwidth]{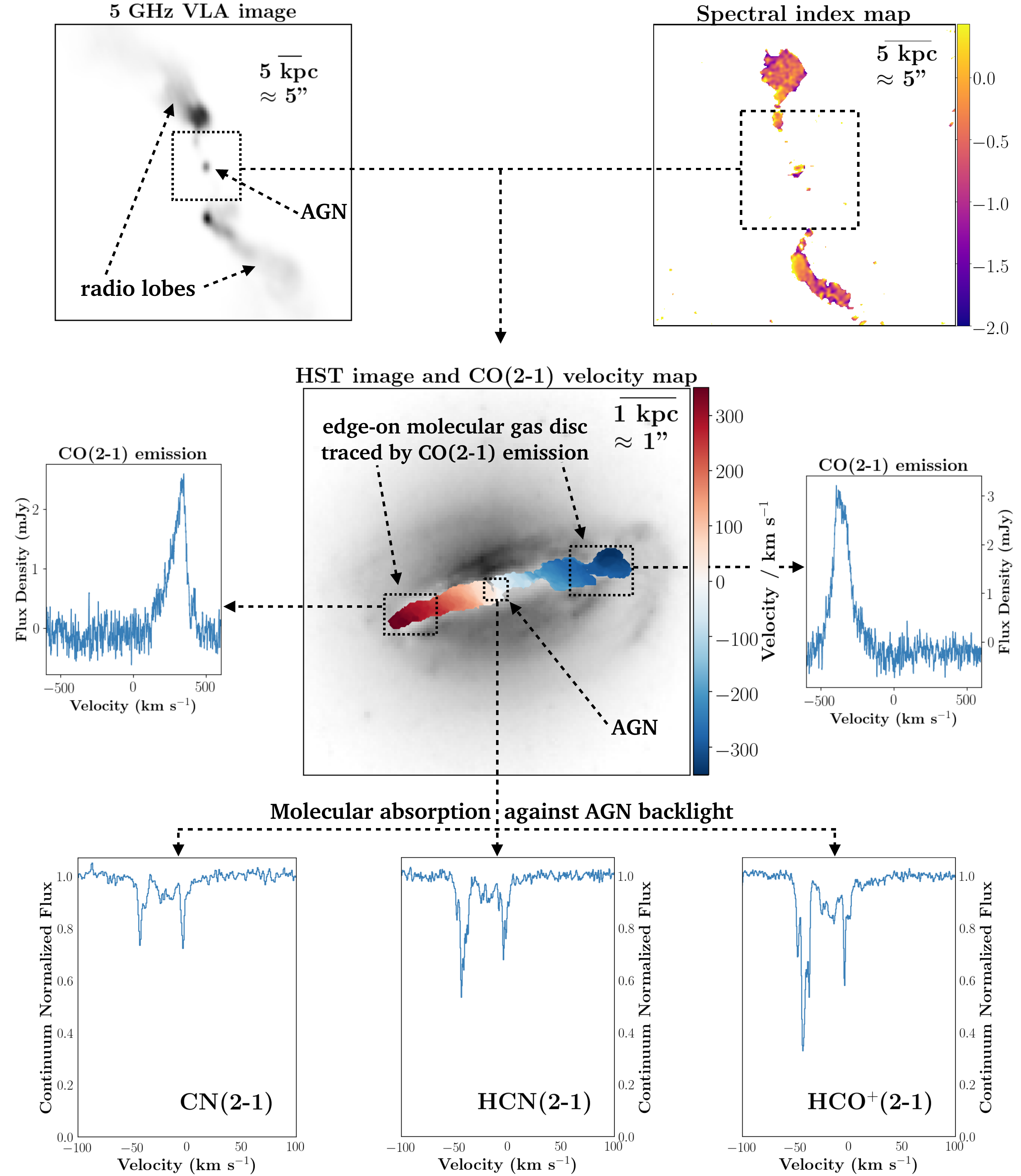}
     \caption{A multi-wavelength view of Hydra-A's AGN, radio lobes and edge-on molecular gas disc. \textbf{Top left:} An unmasked 5 GHz \textit{Karl G. Jansky Very Large Array} (VLA) image showing the galaxy's AGN and its radio lobes emanating to the north and south, with 0.19 arcsec pixel$^{-1}$ resolution (Project 13B-088). \textbf{Top right:} A \mbox{0.29 arcsec pixel$^{-1}$} spectral index map of the AGN and radio lobes, produced from continuum images at 92 and 202 GHz which were taken as part of our ALMA survey. \textbf{Centre:} A \mbox{0.05 arcsec pixel$^{-1}$} F814W \textit{Hubble Space Telescope} (HST) near-infrared image \citep{Mittal2015}. Overlaid is a velocity map which traces the galaxy's edge-on disc of cold molecular gas, produced using our ALMA observations of CO(2-1) emission. \textbf{Centre left and right:} The spectra of CO(2-1) emission from the red and blueshifted sides of the edge-on disc, also extracted from the ALMA data presented in this paper. \textbf{Bottom:} Some of the principal absorption lines seen against the continuum source at the galaxy centre,  which we explore in this paper. The absorption is produced by the cold molecular gas within the disc which lies along the line of sight to the bright radio core.}
    \label{fig:Hydra-A_HST_cont_image}
\end{figure*}

Recent theories and simulations have predicted that supermassive black hole accretion is, to a large extent, powered by the chaotic accretion of clumpy molecular gas clouds \citep[e.g. ][]{Pizzolato2005, voort2012, Gaspari2018}. This accretion is just one element of a galaxy-wide, self-regulating fuelling and feedback cycle \citep{PetersonFabian2006,Voit2015, McNamara2016, Tremblay2018}. The accreted mass powers radio jets, which in turn produce shocks and turbulence throughout the galaxy, as well as inflating buoyant bubbles of hot, X-ray bright gas. Turbulence, rising bubbles and pressure waves produced by these shocks cause localised increases in gas densities, lift clouds to higher altitudes, decrease cooling times and promote the formation of cold molecular gas clouds. The outward velocities of these newly formed clouds of molecular gas are typically much lower than the escape velocity, meaning that significant amounts of this newly formed cold molecular gas eventually returns to the centre of the galaxy to further fuel the feedback loop \citep[for a short review see][]{Gaspari2020}. 

\begin{table*}
	\centering
	\begin{tabular}{lccccr} 

		\hline
		 & CO(1-0) & $^{13}$CO(2-1)& CO(2-1) & HCO$^{+}$(1-0) & HCO$^{+}$(2-1)\\
		 Target lines & & C$^{18}$O(2-1) & CN(2-1) & HCN(1-0) & HCN(2-1)\\
		 &  &  SiO(5-4) & H$_{2}$CO(3-2) & HNC(1-0) & \\
		\hline
		Observation date & 2018 Jul 18 & 2018 Dec 12 & 2018 Oct 30 & 2019 Sep 24 & 2018 Nov 16 \\
		Integration time (mins) & 44 & 215 & 95 & 48 & 85 \\
		Velocity width per channel (km s$^{-1}$)  & 2.7 & 1.4 & 0.7 & 1.7 & 0.9 \\
		Frequency width per channel (kHz) & 977 & 977 & 488 & 488 & 488 \\
		Beam dimensions (") & $2.3 \times 1.6$  & $0.60 \times 0.46$ & $0.27 \times 0.25$ & $0.47\times 0.29$ & $0.38 \times 0.32$ \\
		Spatial resolution (kpc) & 1.71 & 0.54 & 0.29 & 0.44 & 0.36 \\
		Precipitable water vapour (mm)  & 2.85 & 1.59 & 0.96 & 3.21 & 1.04 \\
		Field of view (arcsec)  & 56.9 & 28.9 & 26.1 & 63.3 & 33.4\\
		ALMA band  & 3 & 5 & 6 & 3 & 5 \\
		ALMA configuration & C43-1 & C43-4 & C43-5 & C43-6 & C43-5 \\
		Maximum baseline (m)  & 161 & 784 & 1400 & 2500 & 1400 \\
		Noise/channel (mJy/beam) & 1.01 & 0.27/0.27/0.27 & 1.33/0.47/0.47 & 0.58/0.56/0.58 & 0.57/0.63 \\
		\hline
	\end{tabular}
	\caption{A  summary  of  the  observational details for the ALMA data presented in this paper. Each column of the table represents a different observation, with most containing multiple target lines.}
	\label{tab:observations}
\end{table*}

Observing gas in the cold molecular phase is essential if we are to understand the wider cycle of accretion and feedback. For many decades this has been best achieved with molecular emission line studies \citep[recent examples include][]{GarciaBurillo2014,Temi2018, Ruffa2019, Olivares2019}. However, the emission lines of individual molecular clouds are relatively weak, so studying the molecular gas in this way can only be used to reveal the behaviour of large ensembles of molecular gas clouds. In recent years, several studies have been able to observe molecular gas in the central regions of brightest cluster galaxies through absorption, rather than emission \citep{David2014, Tremblay2016, Ruffa2019, Rose2019a, Rose2019b, Nagai19, Combes2019}. The key advantage of these studies is that they are able to detect the presence of molecular clouds in small groups, or even individually because they make use of a bright central core, against which it is possible to observe molecular absorption along very narrow lines of sight. 

Absorption line studies can be split into two main groups: intervening absorbers and associated absorbers. Intervening absorbers take advantage of chance alignments between galaxies and background quasars, while associated absorbers use the radio source coincident with a galaxy's supermassive black hole as a bright backlight. Associated absorber systems are particularly useful because when using a galaxy's bright radio core as a backlight, redshifted absorption unambiguously indicates inflow and blueshifted lines indicate outflow. In these cases it is possible to make direct observations of gas with knowledge of how it is moving relative to the supermassive black hole and which may even be in the process of accretion, as has been done by \citet{David2014, Tremblay2016, Rose2019b}, where molecular absorption due to clouds moving at hundreds of \kms towards their host supermassive black holes has been detected. From the nine associated absorber systems found in brightest cluster galaxies to date, a tendency has emerged for these absorbing molecular gas clouds to have bulk motions toward the host supermassive black holes \citep{Rose2019b}.

Until recently all absorption line studies in brightest cluster galaxies had searched for carbon monoxide (CO). Although the detection of these systems with CO alone is of great value, observing the same absorption regions with multiple molecular species has the potential to reveal the chemistry and history of the gas in the surroundings of supermassive black holes in much more detail, significantly increasing our understanding of the origins of the gas responsible for their accretion and feedback mechanisms. \citet{Rose2019b} recently presented eight absorbing brightest cluster galaxies, which included seven with CO absorption and seven with low resolution CN absorption. Nevertheless, high spectral resolution observations of these absorption systems with a wider mix of molecular species are still lacking. This paper marks the beginning of a campaign to address this issue.

The observations we present are from an \textit{Atacama Large Millimeter/submillimeter Array} (ALMA) Cycle 6 survey originally designed to detect the absorption lines of several molecular species in Hydra-A, namely CO, $^{13}$CO, C$^{18}$O, CN, HCN and \HCO. A multi-wavelength view of Hydra-A which highlights its main features can be seen in Fig. \ref{fig:Hydra-A_HST_cont_image}. The galaxy is already known to have by far the most optically thick CO absorption of this type, caused by clouds of cold, molecular gas lying along the line of sight to the bright radio source which is spatially coincident with the supermassive black hole \citep{Rose2019a}. These clouds are almost entirely composed of hydrogen, though small but significant amounts of these less common molecules are present at sufficient abundances to produce detectable absorption lines. 

Although no study of a single source can ever be representative of a whole family of astronomical objects, Hydra-A is a prime target for a study of this type for several reasons, perhaps most importantly because it is a giant elliptical galaxy with a near perfectly edge-on disc of dust and molecular gas, which should readily produce absorption lines in the spectrum of any radio source lying behind it \citep{Hamer2014}. Perpendicular to the disc are powerful radio jets and lobes which propagate out of the galaxy's centre and into the surrounding X-ray luminous cluster \citep{Taylor1990}. Over several gigayears the galaxy's AGN outbursts have created multiple cavities in this X-ray emitting gas via the repeated action of these radio jets and lobes \citep{Hansen1995,Hamer2014}. Hydra-A is a particularly useful target for a study of molecular absorption because it is an extremely bright radio source, with one of the highest flux densities in the 3C catalogue of radio sources \citep{Edge1959}. Combined with its compact and unresolved nature, this high flux density makes it an ideal backlight for an absorption line survey. This is particularly true in our case where we have aimed to detect molecular species with relatively low column densities, such as CO isotopologues. Previous observations across several wavelength bands also suggest that the galaxy's core contains a significant mass of both atomic and molecular gas e.g. CO and CN absorption by \citet{Rose2019a, Rose2019b}, H\thinspace\small{I}\normalsize\space absorption by \citet{Dwarakanath1995,Taylor1996}, CO emission by \citet{Hamer2014}, and H$_{2}$ studies by \citet{Edge2002,Donahue2011,Hamer2014}. 

Throughout the paper, velocity corrections applied to the spectra of Hydra-A use a redshift of \mbox{$z$ = 0.0544$\pm$0.0001}, which provides the best estimate of the systemic velocity of the galaxy. This redshift is calculated from MUSE observations of stellar absorption lines (ID: 094.A-0859) and corresponds to a recession velocity of \mbox{16294$\pm$30 km s$^{-1}$}. At this redshift, there is a spatial scale of \mbox{1.056 kpc arcsec$^{-1}$}, meaning that kpc and arcsec scales are approximately equivalent. The CO(2-1) emission line produced by the molecular gas disc also provides a second estimate for the galaxy's recession velocity of 16284 km s$^{-1}$, though this value has a larger uncertainty due to potential gas sloshing.

\section{Observations and target lines}

Observations at the expected frequencies of the CO(1-0), CO(2-1), $^{13}$CO(2-1), C$^{18}$O(2-1), CN(2-1), \HCO(1-0), \HCO(2-1), HCN(1-0), HCN(2-1) and HNC(1-0) rotational lines in Hydra-A were carried out between 2018 July 18 and 2018 Dec 12. The CO(1-0) observation was carried out as part of an ALMA Cycle 4 survey (2016.1.01214.S), and the remaining were part of an ALMA Cycle 6 survey (2018.1.01471.S). Absorption from all of these lines except C$^{18}$O(2-1) was detected. Serendipitous detections of SiO(5-4) and H$_{2}$CO(3-2) were also made during the observations of the target lines. The main details for each observation are given in Table \ref{tab:observations}. For these observations, Figs. \ref{fig:MultiSpectra_Plot_diatomic} and \ref{fig:MultiSpectra_Plot_polyatomic} show the spectra seen against the bright radio source at the centre of the galaxy. All are extracted from a region centered on the continuum source with a size equal to the synthesized beam's FHWM.

With such a wide range of molecular lines targeted, the properties of the gas clouds responsible for the absorption can be revealed in significant detail. A short summary of the particular properties each molecular species can reveal about the absorbing gas clouds is provided below, as well as references to more in depth information for the interested reader. The dipole moments for the molecules observed are given in Table \ref{tab:frequencies_table}, along with the critical density and rest frequency of each line.

\begin{figure*}
	\includegraphics[width=\textwidth]{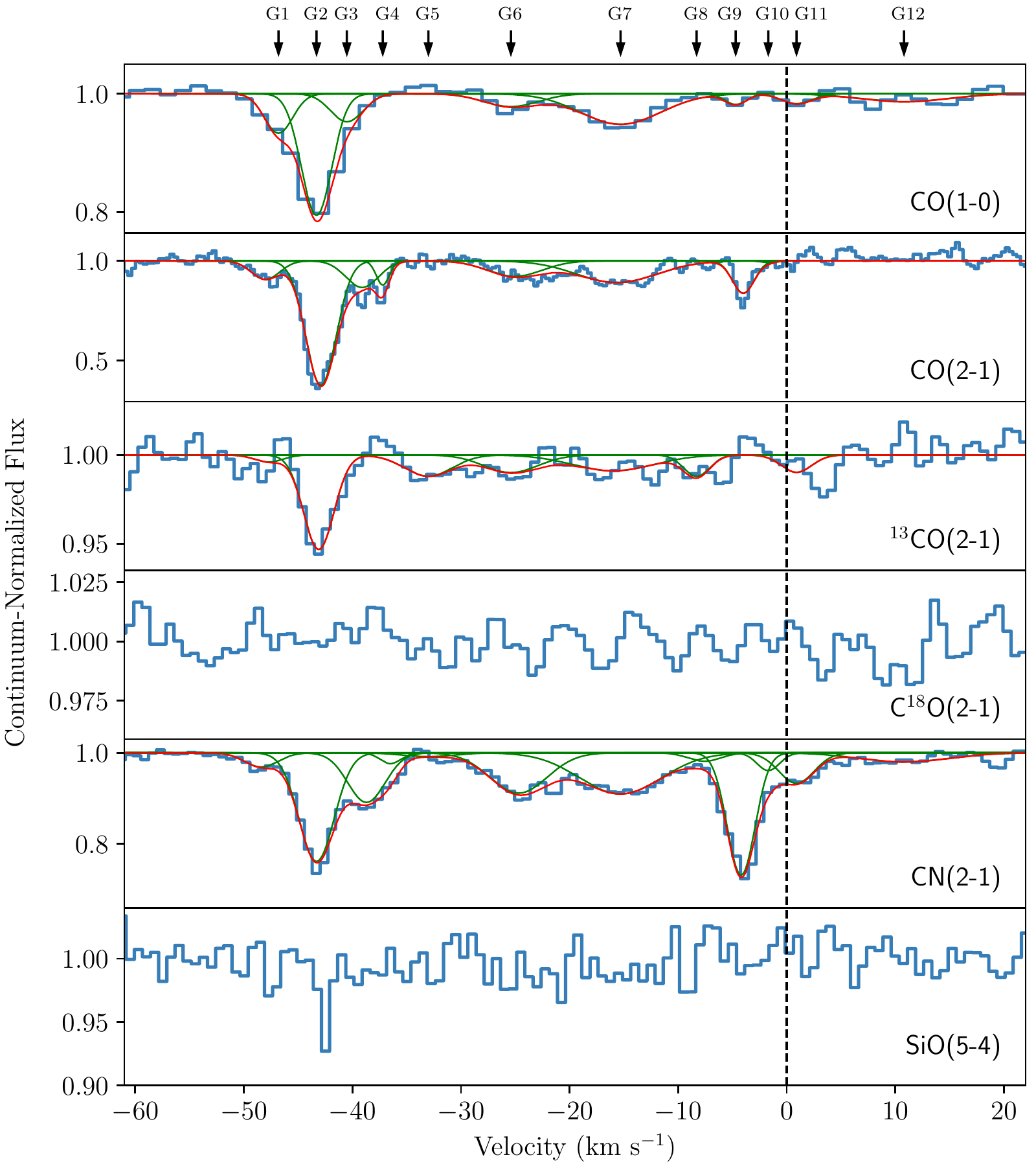}
     \caption{Absorption profiles observed against the bright and compact radio continuum source at the centre of Hydra-A, which is spatially coincident with the brightest cluster galaxy's AGN and supermassive black hole. These spectra have a very narrow velocity range of approximately 80 km s$^{-1}$ in order to show the absorption features clearly. The full width of the observed spectra is typically 2000 km s$^{-1}$, though no absorption features outside the velocity range shown are apparent. The spectra are extracted from a region with a size equal to the FWHM of each observation's synthesized beam. Red and green lines show the individual and combined 12-Gaussian best fits, where each of the 12 Gaussians has a freely varying amplitude across all of the spectra, but a fixed FWHM and central velocity (as indicated by the arrows at the top of the plot). The process by which the best fits are found is described in \S\ref{sec:fitting_procedure}. Some of the 12 Gaussian lines may appear weak and unconvincing in some spectra, but all are resolved and detected to high significance in at least one absorption line. No reliable best fits are found for C$^{18}$O(2-1) or SiO(5-4). Continued in Fig. \ref{fig:MultiSpectra_Plot_polyatomic}.}
    \label{fig:MultiSpectra_Plot_diatomic}
\end{figure*}

\begin{figure*}
	\includegraphics[width=\textwidth]{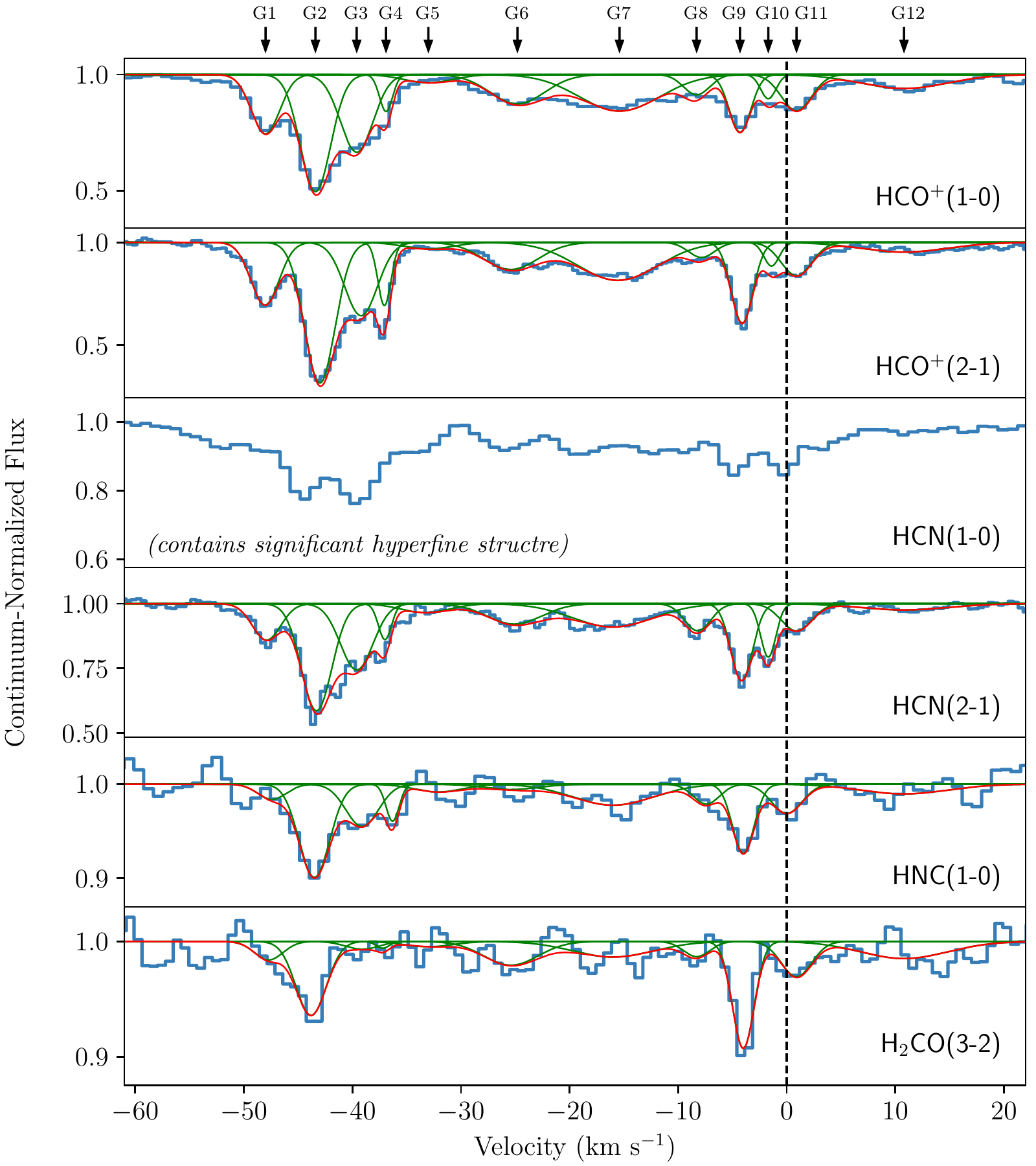}
     \caption{Continued from Fig. \ref{fig:MultiSpectra_Plot_diatomic}. Note that the CN(2-1), HCN(1-0) and HCN(2-1) lines all contain hyperfine structure.}
    \label{fig:MultiSpectra_Plot_polyatomic}
\end{figure*}

\begin{itemize}
    \item \textbf{CO} (carbon monoxide) has a relatively small electric dipole moment which allows it to undergo collisional excitation easily. This makes it readily visible in emission and as a result it is commonly used as a tracer of molecular hydrogen, which has no rotational lines due to its lack of polarization. CO is relatively abundant within the centres of brightest cluster galaxies and has many rotational lines which are sufficiently populated to produce observable emission and absorption lines. The variation in the absorption strengths of these different rotational lines can be used to estimate the excitation temperature of the gas \citep{Magnum2015}. The strength of each absorption line is dependent on the number of CO molecules in each rotational state, which itself is determined by the gas excitation temperature. Therefore, the ratio of the optical depths for various absorption lines of CO can give a direct measure of the gas excitation temperature, assuming that the lines are not optically thick.

    \item \textbf{$^{13}$CO}, when seen at high column densities, is normally associated with galaxy mergers and ultra-luminous infrared galaxies \citep{Taniguchi1999, Glenn2001}, while CO/$^{13}$CO values have been shown to correlate with star formation and top heavy initial mass functions \citep{Davis2014, Sliwa2017}. Variation in the CO/$^{13}$CO ratio is also seen within the Milky Way and other galaxies, with decreasing values associated with proximity to the galaxy centre as a result of astration \citep{Wilson1999, Paglione2001, Vantyghem2017}. $^{13}$CO is typically at least an order of magnitude less abundant than CO, so the absorption lines of this isotopolgue can be used to distinguish between optically thick clouds with a low covering fraction and more diffuse clouds which cover an entire continuum source. For example, if a molecular cloud extinguishes 10 per cent of a continuum source's flux in CO(1-0), it may be an optically thin cloud with $\tau = 0.1$, or an optically thick cloud (i.e. $\tau\gg1$) covering 10 per cent of the continuum. $^{13}$CO(1-0) could distinguish between these scenarios; its absorption would be much more significant and more easily detected in the case of an optically thick cloud.
    
    \item \textbf{C$^{18}$O} contains the stable oxygen-18 isotope, which is predominantly produced in the cores of stars above $8~\textnormal{M}_{\odot}$ \citep{Iben1975}. The ratio of the absorption strength seen from $^{13}$CO, C$^{18}$O, and other CO isotopologues can therefore be used as a probe of the star formation history of the molecular gas in which the molecules are observed \citep[see][]{Papadopoulos1996, Zhang2018, Brown2019}.
    
    \item \textbf{CN} (cyanido radical) molecules are primarily produced by photodissociation reactions of HCN. Its emission lines are therefore normally indicative of molecular gas in the presence of a strong ultraviolet radiation field \citep[for a detailed overview of the origins of CN, see][]{Boger2005}. Models have shown that the production of CN at high column densities can also be induced by the strong X-ray radiation fields found close to AGN \citep{Meijerink2007}. Observations of CN emission lines from nearby galaxies show internal variation in the CO/CN line ratios of around a factor of three \citet{Wilson2018}. System to system variation in the CO/CN ratio of at least an order of magnitude is also observed in the absorption lines of brightest cluster galaxies \citet{Rose2019b}.  
    
    \item \textbf{SiO} (silicon monoxide) is associated with warm, star-forming regions of molecular gas, where it is enhanced by several orders of magnitude compared with darker and colder molecular gas clouds. As a result, SiO is normally linked to dense regions and shocks, though it has occasionally been detected in low density molecular gas via absorption \citep[e.g.][]{Peng1995, Muller2013}.
    
    \item \textbf{\HCO \textnormal{(formyl cation) and} HCN} (hydrogen cyanide) are tracers of low density molecular gas when seen in absorption, since it is only at low densities that the molecules are not collisionally excited to high J-levels. Their absorption lines have been detected in a handful of intervening absorber systems, e.g. \citet{Wiklind1997a, Muller2011}. Due to their large electric dipole moments, the molecules have often been detected with relative ease despite being much less abundant than e.g. CO or CN \citep[e.g.][]{Lucas1996, Liszt2001, Gerin2019, Seiji2020}.
     
     \item \textbf{HNC} (hydrogen isocyanide) is a tautomer of HCN. Thanks to its similar structure, it can be used as a tracer of gas properties in a similar manner to HCN and HCO$^{+}$. HNC detections may also useful in combination with those of HCN because of an observed dependence of the \textit{I}(HCN)/\textit{I}(HCN) ratio on the gas kinetic temperature \citep{Hernandez-Vera2017, Hacar2019}.
     
    \item \textbf{H$_{2}$CO} (formaldehyde) is highly prevalent toward H\thinspace\small II\normalsize\space regions and has been found throughout the interstellar medium at relatively high abundances which do not vary significantly, even in particularly chaotic regions \citep{Henkel1983, Downes1980, Ginard2012}. The molecule has several pathways of formation within the interstellar medium, split into two main groupings. First, it can form on the icy surfaces of dust grains. Second, it can be produced more directly in the gas phase. The formation of H$_{2}$CO on dust grains requires CO to be frozen onto the surface, so this mechanism mainly contributes to H$_{2}$CO gas at distances of hundreds of AU from stars, where temperatures are low enough for volatile molecules to condense \citep{Qi2013, Loomis2015}.
\end{itemize}

\section{Data Processing}
\label{sec:fitting_procedure}

\begin{table*}
	\centering
	\begin{tabular}{lcccr}
		\hline
        Transition & Dipole Moment / $D$ & Critical Density / cm$^{-3}$ & Frequency / GHz & Detected \\
        \hline
        CO(1-0) & 0.112 & $4.1\times 10^2$ & 115.271208 & Yes \\
        CO(2-1) & \textquotedbl & $2.7\times 10^3$ & 230.538000 & Yes \\
        $^{13}$CO(2-1) & 0.112 & $2.7\times 10^3$ &  220.398684 & Yes \\
        C$^{18}$O(2-1) & 0.112 & $2.7\times 10^3$ & 219.560354 & No \\
        CN(2-1) & 1.450 & $1.4\times 10^6$ & 226.874783* & Yes \\
        SiO(5-4) & 3.098 & $1.7\times 10^6$ & 217.104980 & Yes \\
        \HCO(1-0) & 3.300 & $2.3\times 10^4$ & 89.188525 & Yes \\
        \HCO(2-1) & \textquotedbl & $2.2\times 10^5$ & 178.375056 & Yes \\
        HCN(1-0) & 2.980 & $1.1\times 10^5$ & 88.631602* & Yes \\
        HCN(2-1) & \textquotedbl & $1.1\times 10^6$ & 177.261117* & Yes \\
        HNC(1-0) & 3.050 & $7.0\times 10^4$ & 90.663568 & Yes \\
        H$_{2}$CO(3-2) & 2.331 & $4.5\times 10^5$ & 225.697775 & Yes \\
		\hline 
		\multicolumn{4}{c}{*intensity weighted mean of hyperfine structure lines} \\
	\end{tabular}

    \caption{Dipole moments, critical densities and rest frequencies for the molecules discussed in this paper. The critical densities are calculated at kinetic temperatures of 100\thinspace K.}
    \label{tab:frequencies_table}
\end{table*}

\begin{table*}
\centering
\begin{tabular}{lccccc}
\hline
 Line & v$_{\textnormal{cen}}$  / km s$^{-1}$ & $\sigma$ / km s$^{-1}$ & $T_{\textnormal{ex}}$ / K & D / pc & M$_{\textnormal{tot}}$ / $M_{\odot}$ \\
 \hline
G1 & $-47.7$ & 1.3 & 5.1 $^{+ 0.5 }_{- 0.4 }$ & 1.7 & 330 \\
G2 & $-43.1$ & 1.4 & 21.0 $^{+ 54.6 }_{- 7.6 }$ & 2.0 & 450 \\
G3 & $-39.1$ & 1.5 & 4.7 $^{+ 0.3 }_{- 0.3 }$ & 2.3 & 600 \\
G4 & $-37.2$ & 0.6 & 9.7 $^{+ 3.8 }_{- 2.4 }$ & 0.4 & 33 \\
G5 & $-33.0$ & 2.2 & 4.3 $^{+ 4.6 }_{- 1.3 }$ & 4.8 & $2.7\times 10^{3}$ \\
G6 & $-25.4$ & 2.5 & 4.6 $^{+ 0.7 }_{- 0.5 }$ & 6.3 & $4.6 \times 10^{3}$ \\
G7 & $-16.0$ & 3.7 & 4.8 $^{+ 0.3 }_{- 0.3 }$ & 13.7 & $2.2 \times 10^{4}$ \\
G8 & $-8.3$ & 1.2 & 3.4 $^{+ 0.6 }_{- 0.4 }$ & 1.0 & 430 \\
G9 & $-4.0$ & 1.0 & 7.2 $^{+ 1.0 }_{- 0.8 }$ & 1.0 & 430 \\
G10 & $-1.7$ & 0.7 & 4.9 $^{+ 1.8 }_{- 1.0 }$ & 0.5 & 40 \\
G11 & 0.9 & 1.4 & 4.5 $^{+ 0.5 }_{- 0.4 }$ & 2.0 & 450 \\
G12 & 10.8 & 4.0 & 3.5 $^{+ 0.7 }_{- 0.5 }$ & 16.0 & $7.5 \times 10^{3}$\\
\hline
\end{tabular}
\label{tab:FWHM_v_cen_temp_values}
\caption{The central velocities, velocity dispersions, excitation temperatures and corresponding diameters and masses of the absorption regions which make up the 12-Gaussian fit applied to each of the spectra shown in Figs. \ref{fig:MultiSpectra_Plot_diatomic} and \ref{fig:MultiSpectra_Plot_polyatomic}. The fitting procedure by which the central velocities and velocity dispersions are found is described in detail in \S\ref{sec:fitting_procedure}. The excitation temperatures are estimated from the HCO$^{+}$(1-0) and HCO$^{+}$(2-1) lines using using Equation \ref{eq:opacityratio}, while the sizes and masses are found using a size-linewidth relation and with the assumption of virial equilibrium (see \S\ref{sec:size_mass_estimates}).}
\end{table*}

The data presented throughout this paper were handled using CASA version 5.6.0, a software package which is produced and maintained by the National Radio Astronomy Observatory (NRAO) \citep{CASA}. The calibrated data were produced by the ALMA observatory and following their delivery, we made channel maps at maximal spectral resolution. The self-calibration of the images was done as part of the pipeline calibration. 

The values used when converting from the frequencies observed to velocities are given in Table \ref{tab:frequencies_table}. The CN absorption profile in Fig. \ref{fig:MultiSpectra_Plot_diatomic} is composed of three unresolved hyperfine structure lines of the N= 2-1, J=5/2-3/2 transition. We use the intensity weighted mean of these lines as the rest frequency. The full CN(2-1) spectrum, including all of its observed hyperfine structure lines, can be seen in Appendix \ref{sec:CN_HCN_hfs}. HCN(2-1) also contains hyperfine structure, though it is closely spaced enough that it does not significantly affect the appearance of the spectrum. HCN(1-0) contains hyperfine structure at separations which make resolving the 12 absorption regions unfeasible.

\subsection{Line fitting procedure}
Figs. \ref{fig:MultiSpectra_Plot_diatomic} and \ref{fig:MultiSpectra_Plot_polyatomic} show that the relative strengths of the absorption seen in a given velocity range of the spectrum can vary significantly between the molecular tracers. For example, in the CO(2-1) and H$_{2}$CO(3-2) spectra, the absorption features represented by G2 and G9 are the strongest. In CO(2-1) the first is significantly stronger than the second, wheres for H$_{2}$CO(3-2) the reverse is true. The absorption is nevertheless produced by the same two regions of molecular gas, which will have the same velocity dispersion, $\sigma$, and central velocity, $v_{\textnormal{cen}}$, since they are determined by the clouds' gas dynamics and not the abundance of the molecular tracer they are observed with. To reflect this, we find a common multi-Gaussian best fit line which is composed of several Gaussian lines. Each has a fixed $\sigma$ and $v_{\textnormal{cen}}$ across all of the spectra, but a freely varying amplitude. 

To find the minimum number of Gaussian lines needed for a good fit, and their $\sigma$ and $v_{\textnormal{cen}}$, we start with the three best resolved absorption lines: CO(2-1), HCO$^{+}$(2-1) and HCN(2-1). An initial fit was made using 10 Gaussian lines. This is the number which are clearest to the eye on initial inspection of the spectra. In the final fits to the data shown in the plots, these initial 10 are labeled as G1, G2, G3, G4, G6, G7, G8, G9, G10 and G11. G1, G2, G3, G4, G6, G7 are most easily seen in the HCO$^{+}$(2-1) profile, while G8, G9, G10 and G11 are clearest in the HCN(2-1) profile. For the three spectra, best fits are found for a range of $\sigma$ and $v_{\textnormal{cen}}$. The values which provide the lowest reduced $\chi^{2}$ across the three spectra are then used as the basis of the best fit line for all of the spectra. With a fixed $\sigma$ and $v_{\textnormal{cen}}$, the amplitudes of the Gaussian lines are then the only free parameters.

The initial 10 Gaussian fit is found to be insufficient, with $> 5 \sigma$ absorption remaining in the residuals across several neighbouring channels. Two more Gaussians are added to the best fit line (labelled G5 and G12 in the final fit) to account for this extra absorption. Once again, the values of $\sigma$ and $v_{\textnormal{cen}}$ which provide the lowest reduced $\chi^{2}$ across the three spectra are then used as the basis of the best fit line for all of the spectra. The minimum number of Gaussians required to provide a good fit for all of the lines is found to be 12. The $\sigma$ and $v_{\textnormal{cen}}$ of these lines is given in Table \ref{tab:FWHM_v_cen_temp_values}.

It is possible that some of the regions represented by each Gaussian line are made up from absorption due to multiple molecular clouds, rather than an individual one. If this is true a small shift in the central velocity of each Gaussian line across the different molecular transitions may result from any temperature, density or velocity dispersion gradient which exists along the line of sight. This was investigated for each molecular absorption line by allowing the v$_\textnormal{cen}$ of each Gaussian line to vary as a free parameter during the fitting process. The v$_\textnormal{cen}$ values resulting from this process were consistent with the fixed values, so it is not evident that this issue affects the fits shown in Figs. \ref{fig:MultiSpectra_Plot_diatomic} and \ref{fig:MultiSpectra_Plot_polyatomic}.

When applying the final 12 Gaussian fit to all of the spectra, the $\sigma$ are fixed while the $v_{\textnormal{cen}}$ are pinned to a common value, but allowed to vary by up to an amount equal to the spectrum's velocity resolution. The amplitude of each Gaussian is the only free parameter and is able to take any value less than or equal to zero.

To find a final best fit line and errors for the spectra, we use a Monte Carlo approach. For each spectrum the noise was estimated from the root mean square (RMS) of the continuum emission. This was calculated after excluding the region where any emission or absorption is visible. Following this, \mbox{10 000} simulated spectra are created based upon the observed spectrum. To produce each simulated spectrum, a Gaussian distribution is created for each velocity channel. This Gaussian distribution is centred at the intensity in the observed spectrum for that particular velocity channel, and has a variance equal to the RMS noise squared. A random value for the intensity is drawn from the Gaussian distribution and when this has been done across all velocity channels, a simulated spectrum is produced. The fitting procedure described above is then applied to each simulated spectrum to estimate the strength of each of the 12 Gaussian absorption regions. The upper and lower ${1\sigma}$ errors are taken from the values which delimit the 15.865 per cent highest and lowest results for each of the fits (i.e. 68.27 per cent of the fitted parameters will therefore lie within this $1\sigma$ range).

The $\sim~10^{9}\textnormal{M}_{\odot}$ of molecular gas that is present across the disc of the galaxy produces broad CO(1-0) and CO(2-1) emission lines with FWHM of hundreds of \kms \citep[][fig. 2]{Rose2019a}. Since we are primarily interested in the significantly more narrow absorption features which lie at the centre of the emission, the emission is removed from the spectra in the following way. First, a Gaussian fit is made to the emission. During the fitting process, the spectral bins in which absorption can be seen are masked, approximately \mbox{$-50$ to $+10$ \kms}. This masking region was chosen by performing Gaussian fits to the emission after applying masks to the spectra with limits at every spectral bin between \mbox{$-55$ $\pm$ 10 \kms} and \mbox{+5 $\pm$ 10 km s$^{-1}$}. The chosen range produces a spectrum with the lowest \mbox{$\chi^{2}_{\nu}$} value when the non-masked, emission-subtracted region is fitted to a flat line. Corresponding emission lines are not visible in the other molecular species due to their higher electric dipole moments, which makes collisional excitation less likely. The molecular emission lines are therefore so faint as to be undetectable given the integration times of our observations.

\begin{table*}
\centering
\begin{tabular}{llccccccc}
\hline
 &  & G1 & G2 & G3 & G4 & G5 & G6 \\
\hline
\multicolumn{2}{c}{v$_{\textnormal{cen}}$  / km s$^{-1}$} & $-47.7$
& $-43.1$
& $-39.1$
& $-37.2$
& $-33.0$
& $-25.4$ \\
\multicolumn{2}{c}{$\sigma$ / km s$^{-1}$} & 1.3
& 1.4
& 1.5
& 0.6
& 2.2
& 2.5 \\
\hline
CO(1-0) & $\tau_{\textnormal{max}}$ & 0.07 $^{+ 0.02 }_{ -0.02 }$ & 0.35 $^{+ 0.04 }_{ -0.04 }$ & $<0.07$ & $<0.04$ & $<0.01$ & $<0.03$ &  \\
 & $\int~\tau \textnormal{d}v$ / km s$^{-1}$ & 0.10 $^{+ 0.03 }_{ -0.03 }$ & 0.52 $^{+ 0.06 }_{ -0.05 }$ & $<0.1$ & $<0.06$ & $<0.02$ & $<0.05$ &  \\
 & N / $\times10^{15}$cm$^{-2}$ & 0.2 $_{- 0.1 }^{+ 0.1 }$ & 13.1 $_{- 1.3 }^{+ 1.5 }$ & $<0.2$ & $<0.4$ & $<0.1$ & $<0.2$ &  \\
 &  &  &  &  &  &  &  &  \\
CO(2-1) & $\tau_{\textnormal{max}}$ & 0.10 $^{+ 0.01 }_{ -0.01 }$ & - & 0.15 $^{+ 0.01 }_{ -0.01 }$ & 0.13 $^{+ 0.02 }_{ -0.02 }$ & $<0.02$ & 0.086 $^{+ 0.007 }_{ -0.008 }$ &  \\
 & $\int~\tau \textnormal{d}v$ / km s$^{-1}$ & 0.15 $^{+ 0.02 }_{ -0.02 }$ & - & 0.22 $^{+ 0.02 }_{ -0.02 }$ & 0.20 $^{+ 0.03 }_{ -0.03 }$ & $<0.03$ & 0.13 $^{+ 0.01 }_{ -0.01 }$ &  \\
 & N / $\times10^{14}$cm$^{-2}$ & 4.0 $_{- 0.5 }^{+ 0.3 }$ & - & 5.4 $_{- 0.4 }^{+ 0.4 }$ & 11.8 $_{- 1.2 }^{+ 1.8 }$ & $<0.5$ & 3.0 $_{- 0.2 }^{+ 0.3 }$ &  \\
 &  &  &  &  &  &  &  &  \\
$^{13}$CO(2-1) & $\tau_{\textnormal{max}}$ & $<0.03$ & 0.07 $^{+ 0.01 }_{ -0.02 }$ & $<0.02$ & $<0.01$ & $<0.02$ & $<0.02$ &  \\
 & $\int~\tau \textnormal{d}v$ / km s$^{-1}$ & $<0.04$ & 0.103 $^{+ 0.022 }_{ -0.022 }$ & $<0.01$ & $<0.01$ & $<0.03$ & $<0.03$ &  \\
 & N / $\times10^{13}$cm$^{-2}$ & $<9.0$ & 224.5 $_{- 54.5 }^{+ 41.4 }$ & $<3$ & $<1$ & $<7$ & $<8$ &  \\
 &  &  &  &  &  &  &  &  \\
CN(2-1)$^{*}$ & $\tau_{\textnormal{max}}$ & 0.03 $^{+ 0.01 }_{ -0.01 }$ & 0.43 $^{+ 0.02 }_{ -0.02 }$ & 0.11 $^{+ 0.01 }_{ -0.01 }$ & 0.02 $^{+ 0.01 }_{ -0.01 }$ & $<0.03$ & 0.09 $^{+ 0.01 }_{ -0.01 }$ &  \\
 & $\int~\tau \textnormal{d}v$ / km s$^{-1}$ & 0.04 $^{+ 0.01 }_{ -0.01 }$ & 0.64 $^{+ 0.03 }_{ -0.03 }$ & 0.17 $^{+ 0.02 }_{ -0.02 }$ & 0.04 $^{+ 0.02 }_{ -0.02 }$ & $<0.04$ & 0.14 $^{+ 0.01 }_{ -0.01 }$ &  \\
 & N / $\times10^{12}$cm$^{-2}$ & 1.0 $_{- 0.3 }^{+ 0.3 }$ & 159 $_{- 8 }^{+ 8 }$ & 4.1 $_{- 0.5 }^{+ 0.5 }$ & 2.3 $_{- 1.2 }^{+ 1.2 }$ & $<0.8$ & 3.3 $_{- 0.2 }^{+ 0.2 }$ &  \\
 &  &  &  &  &  &  &  &  \\
HCO$^{+}$(1-0) & $\tau_{\textnormal{max}}$ & 0.29 $^{+ 0.02 }_{ -0.02 }$ & 1.29 $^{+ 0.07 }_{ -0.06 }$ & 0.40 $^{+ 0.02 }_{ -0.02 }$ & 0.18 $^{+ 0.02 }_{ -0.02 }$ & 0.04 $^{+ 0.01 }_{ -0.01 }$ & 0.13 $^{+ 0.01 }_{ -0.01 }$ &  \\
 & $\int~\tau \textnormal{d}v$ / km s$^{-1}$ & 0.44 $^{+ 0.03 }_{ -0.03 }$ & 1.9 $^{+ 0.1 }_{ -0.1 }$ & 0.59 $^{+ 0.03 }_{ -0.03 }$ & 0.26 $^{+ 0.04 }_{ -0.03 }$ & 0.05 $^{+ 0.02 }_{ -0.02 }$ & 0.19 $^{+ 0.02 }_{ -0.02 }$ &  \\
 & N / $\times10^{12}$cm$^{-2}$ & 1.8 $_{- 0.1 }^{+ 0.1 }$ & 310 $_{- 20 }^{+ 20 }$ & 2.3 $_{- 0.1 }^{+ 0.1 }$ & 3.1 $_{- 0.4 }^{+ 0.5 }$ & 0.1 $_{- 0.1 }^{+ 0.1 }$ & 0.6 $_{- 0.1 }^{+ 0.1 }$ &  \\
 &  &  &  &  &  &  &  &  \\
HCO$^{+}$(2-1) & $\tau_{\textnormal{max}}$ & 0.36 $^{+ 0.01 }_{ -0.01 }$ & - & 0.44 $^{+ 0.01 }_{ -0.01 }$ & 0.37 $^{+ 0.02 }_{ -0.02 }$ & 0.03 $^{+ 0.01 }_{ -0.01 }$ & 0.14 $^{+ 0.01 }_{ -0.01 }$ &  \\
 & $\int~\tau \textnormal{d}v$ / km s$^{-1}$ & 0.54 $^{+ 0.02 }_{ -0.02 }$ & - & 0.66 $^{+ 0.02 }_{ -0.02 }$ & 0.55 $^{+ 0.03 }_{ -0.03 }$ & 0.05 $^{+ 0.01 }_{ -0.01 }$ & 0.21 $^{+ 0.01 }_{ -0.01 }$ &  \\
 & N / $\times10^{12}$cm$^{-2}$ & 2.4 $_{- 0.1 }^{+ 0.1 }$ & - & 2.7 $_{- 0.1 }^{+ 0.1 }$ & 6.0 $_{- 0.3 }^{+ 0.3 }$ & 0.2 $_{- 0.1 }^{+ 0.1 }$ & 0.7 $_{- 0.1 }^{+ 0.1 }$ &  \\
 &  &  &  &  &  &  &  &  \\
HCN(2-1) & $\tau_{\textnormal{max}}$ & 0.15 $^{+ 0.02 }_{ -0.02 }$ & 0.88 $^{+ 0.06 }_{ -0.06 }$ & 0.30 $^{+ 0.02 }_{ -0.02 }$ & 0.15 $^{+ 0.03 }_{ -0.03 }$ & 0.03 $^{+ 0.01 }_{ -0.01 }$ & 0.08 $^{+ 0.01 }_{ -0.01 }$ &  \\
 & $\int~\tau \textnormal{d}v$ / km s$^{-1}$ & 0.22 $^{+ 0.03 }_{ -0.03 }$ & 1.31 $^{+ 0.09 }_{ -0.09 }$ & 0.45 $^{+ 0.04 }_{ -0.04 }$ & 0.23 $^{+ 0.05 }_{ -0.05 }$ & 0.05 $^{+ 0.02 }_{ -0.02 }$ & 0.12 $^{+ 0.02 }_{ -0.02 }$ &  \\
 & N / $\times10^{12}$cm$^{-2}$ & 1.1 $_{- 0.2 }^{+ 0.2 }$ & 84.6 $_{- 5.8 }^{+ 5.8 }$ & 2.2 $_{- 0.2 }^{+ 0.2 }$ & 2.9 $_{- 0.6 }^{+ 0.6 }$ & 0.2 $_{- 0.1 }^{+ 0.1 }$ & 0.5 $_{- 0.1 }^{+ 0.1 }$ &  \\
 &  &  &  &  &  &  &  &  \\
HNC(1-0) & $\tau_{\textnormal{max}}$ & 0.02 $^{+ 0.01 }_{ -0.01 }$ & 0.15 $^{+ 0.02 }_{ -0.02 }$ & 0.04 $^{+ 0.01 }_{ -0.01 }$ & 0.02 $^{+ 0.02 }_{ -0.02 }$ & 0.01 $^{+ 0.01 }_{ -0.01 }$ & 0.01 $^{+ 0.01 }_{ -0.01 }$ &  \\
 & $\int~\tau \textnormal{d}v$ / km s$^{-1}$ & 0.02 $^{+ 0.02 }_{ -0.02 }$ & 0.23 $^{+ 0.03 }_{ -0.03 }$ & 0.07 $^{+ 0.02 }_{ -0.02 }$ & 0.03 $^{+ 0.03 }_{ -0.03 }$ & 0.02 $^{+ 0.01 }_{ -0.01 }$ & 0.01 $^{+ 0.01 }_{ -0.01 }$ &  \\
 & N / $\times10^{11}$cm$^{-2}$ & 0.8 $_{- 0.8 }^{+ 0.8 }$ & 114.6 $_{- 14.9 }^{+ 14.9 }$ & 2.6 $_{- 0.7 }^{+ 0.7 }$ & 3.5 $_{- 3.5 }^{+ 3.5 }$ & 0.6 $_{- 0.3 }^{+ 0.3 }$ & 0.4 $_{- 0.4 }^{+ 0.4 }$ &  \\
 &  &  &  &  &  &  &  &  \\
H$_{2}$CO(3-2) & $\tau_{\textnormal{max}}$ & $<0.02$ & 0.10 $^{+ 0.01 }_{ -0.01 }$ & $<0.02$ & $<0.02$ & $<0.02$ & $<0.01$ &  \\
 & $\int~\tau \textnormal{d}v$ / km s$^{-1}$ & $<0.03$ & 0.15 $^{+ 0.01 }_{ -0.01 }$ & $<0.04$ & $<0.03$ & $<0.02$ & $<0.02$ &  \\
 & N / $\times10^{13}$cm$^{-2}$ & $<4$ & 55 $_{- 3 }^{+ 6 }$ & $<6$ & $<9$ & $<4$ & $<7$ &  \\
\hline
\end{tabular}
\caption{The peak optical depths, velocity integrated optical depths and line of sight column densities for the 12-Gaussian fits applied to each of the spectra shown in Figs. \ref{fig:MultiSpectra_Plot_diatomic} and \ref{fig:MultiSpectra_Plot_polyatomic}. A fit composed of 12 individual Gaussian lines (labelled G1 to G12) of fixed v$_{\textnormal{cen}}$ and $\sigma$, but varying amplitude, is used when fitting to the spectra. Column densities for G2 could not always be reliably calculated because it is optically thick in some of the lines. Continued in Table \ref{tab:mc_fits_tableb}. \newline *The values for CN(2-1) are calculated from three overlapping hyperfine structure lines representing $\sim 60$ per cent of the total absorption. The full CN(2-1) spectrum is shown in Appendix \ref{sec:CN_HCN_hfs}. }
\label{tab:mc_fits_tablea}
\end{table*}

\begin{table*}
\centering
\begin{tabular}{llccccccc}
\hline
 & &  G7 & G8 & G9 & G10 & G11 & G12 \\
\hline
\multicolumn{2}{c}{v$_{\textnormal{cen}}$  / km s$^{-1}$}
& $-16.0$
& $-8.3$
& $-4.0$
& $-1.7$
& 0.9
& 10.8 \\
\multicolumn{2}{c}{$\sigma$ / km s$^{-1}$}
& 3.7
& 1.2
& 1.0
& 0.7
& 1.4
& 4.0 \\
\hline
CO(1-0) & $\tau_{\textnormal{max}}$ & 0.05 $^{+ 0.01 }_{ -0.01 }$ & $<0.02$ & $<0.06$ & $<0.04$ & $<0.06$ & $<0.03$ &  \\
 & $\int~\tau \textnormal{d}v$ / km s$^{-1}$ & 0.08 $^{+ 0.02 }_{ -0.02 }$ & $<0.02$ & $<0.06$ & $<0.06$ & $<0.08$ & $<0.04$ &  \\
 & N / $\times10^{15}$cm$^{-2}$ & 0.20 $_{- 0.04 }^{+ 0.04 }$ & $<0.1$ & $<0.3$ & $<0.3$ & $<0.3$ & $<0.2$ &  \\
 &  &  &  &  &  &  &  &  \\
CO(2-1) & $\tau_{\textnormal{max}}$ & 0.12 $^{+ 0.01 }_{ -0.01 }$ & $<0.03$ & 0.18 $^{+ 0.01 }_{ -0.01 }$ & $<0.03$ & $<0.02$ & $<0.01$ &  \\
 & $\int~\tau \textnormal{d}v$ / km s$^{-1}$ & 0.18 $^{+ 0.01 }_{ -0.01 }$ & $<0.05$ & 0.27 $^{+ 0.02 }_{ -0.02 }$ & $<0.06$ & $<0.03$ & $<0.02$ &  \\
 & N / $\times10^{14}$cm$^{-2}$ & 4.4 $_{- 0.2 }^{+ 0.3 }$ & $<0.5$ & 11.3 $_{- 0.9 }^{+ 0.7 }$ & $<1$ & $<0.6$ & $<0.2$ &  \\
 &  &  &  &  &  &  &  &  \\
$^{13}$CO(2-1) & $\tau_{\textnormal{max}}$ & $<0.01$ & $<0.3$ & $<0.01$ & $<0.01$ & $<0.02$ & $<0.01$ &  \\
 & $\int~\tau \textnormal{d}v$ / km s$^{-1}$ & $<0.02$ & $<0.04$ & $<0.007$ & $<0.01$ & $<0.03$ & $<0.02$ &  \\
 & N / $\times10^{13}$cm$^{-2}$ & $<3$ & $<7$ & $<2$ & $<3$ & $<10$ & $<3$ &  \\
 &  &  &  &  &  &  &  &  \\
CN(2-1)$^{*}$ & $\tau_{\textnormal{max}}$ & 0.10 $^{+ 0.01 }_{ -0.01 }$ & 0.03 $^{+ 0.01 }_{ -0.01 }$ & 0.32 $^{+ 0.02 }_{ -0.02 }$ & 0.04 $^{+ 0.01 }_{ -0.01 }$ & 0.07 $^{+ 0.01 }_{ -0.01 }$ & 0.02 $^{+ 0.01 }_{ -0.01 }$ &  \\
 & $\int~\tau \textnormal{d}v$ / km s$^{-1}$ & 0.14 $^{+ 0.01 }_{ -0.01 }$ & 0.04 $^{+ 0.02 }_{ -0.02 }$ & 0.48 $^{+ 0.02 }_{ -0.02 }$ & 0.06 $^{+ 0.02 }_{ -0.02 }$ & 0.11 $^{+ 0.02 }_{ -0.02 }$ & 0.03 $^{+ 0.01 }_{ -0.01 }$ &  \\
 & N / $\times10^{12}$cm$^{-2}$ & 3.5 $_{- 0.2 }^{+ 0.2 }$ & 0.7 $_{- 0.3 }^{+ 0.3 }$ & 22.0 $_{- 0.9 }^{+ 0.9 }$ & 1.5 $_{- 0.5 }^{+ 0.5 }$ & 2.5 $_{- 0.5 }^{+ 0.5 }$ & 0.5 $_{- 0.2 }^{+ 0.2 }$ &  \\
 &  &  &  &  &  &  &  &  \\
HCO$^{+}$(1-0) & $\tau_{\textnormal{max}}$ & 0.18 $^{+ 0.01 }_{ -0.01 }$ & 0.10 $^{+ 0.02 }_{ -0.02 }$ & 0.29 $^{+ 0.02 }_{ -0.02 }$ & 0.11 $^{+ 0.02 }_{ -0.02 }$ & 0.17 $^{+ 0.02 }_{ -0.02 }$ & 0.06 $^{+ 0.01 }_{ -0.01 }$ &  \\
 & $\int~\tau \textnormal{d}v$ / km s$^{-1}$ & 0.26 $^{+ 0.01 }_{ -0.01 }$ & 0.15 $^{+ 0.02 }_{ -0.02 }$ & 0.43 $^{+ 0.03 }_{ -0.03 }$ & 0.16 $^{+ 0.03 }_{ -0.03 }$ & 0.25 $^{+ 0.02 }_{ -0.02 }$ & 0.09 $^{+ 0.01 }_{ -0.01 }$ &  \\
 & N / $\times10^{12}$cm$^{-2}$ & 0.90 $_{- 0.01 }^{+ 0.01 }$ & 0.30 $_{- 0.01 }^{+ 0.01 }$ & 3.4 $_{- 0.2 }^{+ 0.2 }$ & 0.6 $_{- 0.1 }^{+ 0.1 }$ & 0.8 $_{- 0.1 }^{+ 0.1 }$ & 0.20 $_{- 0.01 }^{+ 0.01 }$ &  \\
 &  &  &  &  &  &  &  &  \\
HCO$^{+}$(2-1) & $\tau_{\textnormal{max}}$ & 0.2 $^{+ 0.01 }_{ -0.01 }$ & 0.07 $^{+ 0.01 }_{ -0.01 }$ & 0.49 $^{+ 0.02 }_{ -0.02 }$ & 0.13 $^{+ 0.01 }_{ -0.01 }$ & 0.18 $^{+ 0.01 }_{ -0.01 }$ & 0.05 $^{+ 0.01 }_{ -0.0 }$ &  \\
 & $\int~\tau \textnormal{d}v$ / km s$^{-1}$ & 0.3 $^{+ 0.01 }_{ -0.01 }$ & 0.11 $^{+ 0.01 }_{ -0.01 }$ & 0.74 $^{+ 0.02 }_{ -0.02 }$ & 0.19 $^{+ 0.02 }_{ -0.02 }$ & 0.27 $^{+ 0.01 }_{ -0.01 }$ & 0.07 $^{+ 0.01 }_{ -0.01 }$ &  \\
 & N / $\times10^{12}$cm$^{-2}$ & 1.10 $_{- 0.01 }^{+ 0.01 }$ & 0.30 $_{- 0.01 }^{+ 0.01 }$ & 5.6 $_{- 0.2 }^{+ 0.2 }$ & 0.7 $_{- 0.1 }^{+ 0.1 }$ & 1.00 $_{- 0.01 }^{+ 0.01 }$ & 0.20 $_{- 0.01 }^{+ 0.01 }$ &  \\
 &  &  &  &  &  &  &  &  \\
HCN(2-1) & $\tau_{\textnormal{max}}$ & 0.10 $^{+ 0.01 }_{ -0.01 }$ & 0.10 $^{+ 0.02 }_{ -0.02 }$ & 0.34 $^{+ 0.03 }_{ -0.03 }$ & 0.21 $^{+ 0.03 }_{ -0.03 }$ & 0.12 $^{+ 0.02 }_{ -0.02 }$ & $<0.04$ &  \\
 & $\int~\tau \textnormal{d}v$ / km s$^{-1}$ & 0.14 $^{+ 0.02 }_{ -0.02 }$ & 0.15 $^{+ 0.03 }_{ -0.03 }$ & 0.51 $^{+ 0.05 }_{ -0.04 }$ & 0.31 $^{+ 0.05 }_{ -0.05 }$ & 0.18 $^{+ 0.03 }_{ -0.03 }$ & $<0.08$ &  \\
 & N / $\times10^{12}$cm$^{-2}$ & 0.6 $_{- 0.1 }^{+ 0.1 }$ & 0.5 $_{- 0.1 }^{+ 0.1 }$ & 4.6 $_{- 0.4 }^{+ 0.4 }$ & 1.5 $_{- 0.2 }^{+ 0.2 }$ & 0.8 $_{- 0.1 }^{+ 0.1 }$ & $<0.3$ &  \\
 &  &  &  &  &  &  &  &  \\
HNC(1-0) & $\tau_{\textnormal{max}}$ & 0.02 $^{+ 0.01 }_{ -0.01 }$ & 0.02 $^{+ 0.01 }_{ -0.01 }$ & 0.08 $^{+ 0.02 }_{ -0.02 }$ & $<0.04$ & 0.03 $^{+ 0.01 }_{ -0.01 }$ & $<0.03$ &  \\
 & $\int~\tau \textnormal{d}v$ / km s$^{-1}$ & 0.03 $^{+ 0.01 }_{ -0.01 }$ & 0.03 $^{+ 0.02 }_{ -0.02 }$ & 0.11 $^{+ 0.02 }_{ -0.02 }$ & $<0.04$ & 0.04 $^{+ 0.02 }_{ -0.02 }$ & $<0.04$ &  \\
 & N / $\times10^{11}$cm$^{-2}$ & 1.1 $_{- 0.4 }^{+ 0.4 }$ & 0.7 $_{- 0.5 }^{+ 0.5 }$ & 8.1 $_{- 1.5 }^{+ 1.5 }$ & $<2$ & 1.4 $_{- 0.7 }^{+ 0.7 }$ & $<0.7$ &  \\
 &  &  &  &  &  &  &  &  \\
H$_{2}$CO(3-2) & $\tau_{\textnormal{max}}$ & 0.02 $^{+ 0.01 }_{ -0.01 }$ & 0.02 $^{+ 0.01 }_{ -0.01 }$ & 0.10 $^{+ 0.01 }_{ -0.01 }$ & $<0.02$ & 0.03 $^{+ 0.01 }_{ -0.01 }$ & 0.02 $^{+ 0.01 }_{ -0.01 }$ &  \\
 & $\int~\tau \textnormal{d}v$ / km s$^{-1}$ & 0.02 $^{+ 0.01 }_{ -0.01 }$ & 0.02 $^{+ 0.01 }_{ -0.01 }$ & 0.16 $^{+ 0.01 }_{ -0.01 }$ & $<0.02$ & 0.05 $^{+ 0.01 }_{ -0.01 }$ & 0.02 $^{+ 0.01 }_{ -0.01 }$ &  \\
 & N / $\times10^{13}$cm$^{-2}$ & 3.7 $_{- 1.4 }^{+ 0.2 }$ & 3.2 $_{- 1.7 }^{+ 0.8 }$ & 29 $_{- 3 }^{+ 2 }$ & $<2$ & 7.0 $_{- 0.8 }^{+ 2.0 }$ & 3.3 $_{- 1.1 }^{+ 0.3 }$ &  \\
\hline
\end{tabular}
\caption{Continued from Table \ref{tab:mc_fits_tablea}.}
\label{tab:mc_fits_tableb}
\end{table*}

\subsection{Optical depth calculations}
The apparent optical depth of an absorption line, $\tau$, can be derived according to the equation,
\begin{equation}
    \tau = -\ln \left( 1 - \frac{1}{f_{\textnormal{c}}}\frac{I_{\textnormal{obs}}}{I_{\textnormal{cont}}} \right),
\end{equation}{}
where $f_{\textnormal{c}}$ is the fraction of the background continuum source covered by the absorbing molecular cloud, $I_{\textnormal{obs}}$ is the depth of the absorption, and $I_{\textnormal{cont}}$ is the continuum level.

We assume a covering factor of 0.7 for the G2 absorption feature at $-43.1$ km s$^{-1}$. Simply assuming $f_{\textnormal{c}}=1$ gives a relatively high $^{13}$CO(2-1) optical depth of $\tau = 0.07$, so for the significantly more ubiquitous CO(2-1), we would expect $\tau\gg 1$ and for the continuum normalized flux to drop to 0. In fact, the line flattens out when around 30 percent of the continuum can still be seen, despite being covered by an optically thick cloud. This in turn implies that the G2 feature covers around 70 per cent of the continuum source.

No highly significant $^{13}\textnormal{CO}$(2-1) absorption is detected in the rest of the absorption profile, as would be expected in the case of optically thick clouds which cover a small fraction of the continuum. Hence, we assume a covering factor of $f_{\textnormal{c}} = 1$ for the remaining absorption features. It is nevertheless possible that we are observing optically thin clouds which do not cover the entire continuum source, so our estimates of $\tau$ are essentially lower limits. Additionally, it is generally assumed that as frequency increases, the emission from an AGN originates closer to its core, so the covering factor may also increase with frequency.

For each spectrum, the implied optical depths of the 12 Gaussian regions are given in Tables \ref{tab:mc_fits_tablea} and \ref{tab:mc_fits_tableb}. The tables also give their velocity integrated optical depths and the implied line of sight column densities, the calculations of which are described in a later section.

\section{Temperature Estimates}
\subsection{Excitation temperature estimates}
The absorption profiles seen in Figs. \ref{fig:MultiSpectra_Plot_diatomic} and \ref{fig:MultiSpectra_Plot_polyatomic} are produced by what we find is best described as the combination of 12 Gaussian absorption regions. Most of the absorption regions have extremely narrow velocity dispersions of $\sim 1~\textnormal{km s}^{-1}$, which are comparable to those of individual clouds in the Milky Way \citep{RomanDuval2010}. Therefore, most of the absorption regions detected can be approximated as individual molecular gas clouds, for which the excitation temperature can be estimated. Even for the broader absorption regions (G7 and G12), which are likely small associations of clouds, an average excitation temperature can still be found. We stress that this concept of individual molecular clouds is an approximation given that there is no clear point where they will start and end, they will have internal structure, and there will always be some interstellar medium which exists between them.

\label{sec:temp_from_HCO}
Our observations of \HCO(1-0) and \HCO(2-1) provide two well resolved absorption profiles from which it is possible to estimate the excitation temperature of the absorption regions represented by each of the 12 Gaussian best fit lines. This requires that the gas is optically thin and in local thermodynamic equilibrium, but as we show in \S\ref{sec:kinetic_temps} this is not the case, so the values should only be treated as approximations.

Nevertheless, with this assumption the \HCO(1-0) and \HCO(2-1) velocity integrated optical depths are related by
\begin{equation}
\label{eq:opacityratio}
\frac{\int \tau_{21} dv}{\int \tau_{10} dv} = 2 \frac{1 - \exp({- h\nu_{21}/k T_{\textnormal{ex}}})}{\exp({h\nu_{10}/k T_{\textnormal{ex}})} -1}\enspace ,
\end{equation}
where $h$ and $k$ are the Planck and Boltzmann constants, $\nu_{10}$ and $\nu_{21}$ are the rest frequencies of the \HCO(1-0) and \HCO(2-1) lines and $T_{\textnormal{ex}}$ is the excitation temperature \citep{Bolatto2003,Godard2010,Magnum2015}. The excitation temperatures found using Equation \ref{eq:opacityratio} are given in Table \ref{tab:FWHM_v_cen_temp_values}.

The hyperfine structure components of the HCN(1-0) line are separated by frequencies similar to those of the 12 Gaussian absorption lines seen in this system. It is therefore not possible to estimate the excitation temperature of the individual absorption regions using the HCN(1-0) and HCN(2-1) spectra, though an average can be estimated from the total velocity integrated optical depth of the whole absorption profile. This gives an excitation temperature of 5.5$^{+ 2.0 }_{- 1.6 }$ K, which compares well with the value of 5.8$^{+ 0.7 }_{- 0.7}$ K when the same method is applied to the HCO$^{+}$(1-0) and HCO$^{+}$(2-1) spectra.

\subsection{Kinetic temperature estimate}
\label{sec:kinetic_temps}
The relative abundance of the HCN and HNC tautomers is observed to depend upon the gas kinetic temperature, with the ratio HCN/HNC increasing at higher temperatures due to reactions which preferentially destroy the HNC molecule \citep{Hernandez-Vera2017, Hacar2019}. Where the intensity ratio satisfies \textit{I}(HCN)/\textit{I}(HCN) $\leq 4$, it is found to correlate with kinetic temperature according to:
\begin{equation}
    {T_\textnormal{kin}} = 10 \times \left[\frac{ I\textnormal{(HCN)}}{I\textnormal{(HCN)}}\right].
\label{eq:HCN_HNC_temp}
\end{equation}{}

Due to the HCN(1-0) line's hyperfine structure, it is only possible to estimate an average kinetic temperature for the absorption profile as a whole, which we find to be $33^{+9}_{-8}$K. The errors quoted are determined from the combination of the uncertainty in the velocity integrated intensities of both spectra and the uncertainty given by \citet{Hacar2019} for Eq. \ref{eq:HCN_HNC_temp}. Since $T_{\textnormal{ex}}< T_{\textnormal{kin}}$, the absorbing gas is sub-thermally excited i.e. it is not in thermal equilibrium.

\section{Column Density Estimates}

The total line of sight column density, $N_{\textnormal{tot}}$, of the absorption regions can be found by using an estimated excitation temperature and assuming that the absorption is optically thin. In general,
\begin{equation}
\label{}
N_{\textnormal{tot}}^{\textnormal{thin}} = Q(T_{\textnormal{ex}}) \frac{8 \pi \nu_{ul}^{3}}{c^{3}}\frac{g_{l}}{g_{u}}\frac{1}{A_{ul}}\frac{1}{R} \frac{1}{ 1 - e^{-h\nu_{ul}/k T_{\textnormal{ex}}}}\int \tau_{ul}~ dv ~,
\label{eq:thin_colum_density}
\end{equation}
where $Q$($T_{\textnormal{ex}}$) is the partition function, $c$ is the speed of light, $A_{ul}$ is the Einstein coefficient of the observed transition and $g$ the level degeneracy, with the subscripts $u$ and $l$ representing the upper and lower levels \citep{Godard2010,Magnum2015}. The factor $R$ is the total intensity of the hyperfine structure lines in the absorption profile, where the combined intensity of all hyperfine lines is normalized to 1. As previously stated, this calculation assumes that the absorption is optically thin. However, in some cases where $\tau \gtrapprox 1$ e.g. G2 of CO(2-1), \HCO(2-1) and HCN(2-1), the true column densities may be significantly higher than calculated. We therefore apply an optical depth correction factor from \citet{Magnum2015} to give a more accurate value for the line of sight column densities,
\begin{equation}
N_{\textnormal{tot}} = N_{\textnormal{tot}}^{\textnormal{thin}} \frac{\tau}{1-\exp({-\tau})}.
\label{eq:true_column_densities}
\end{equation}{}

\begin{figure*}
	\includegraphics[width=\textwidth]{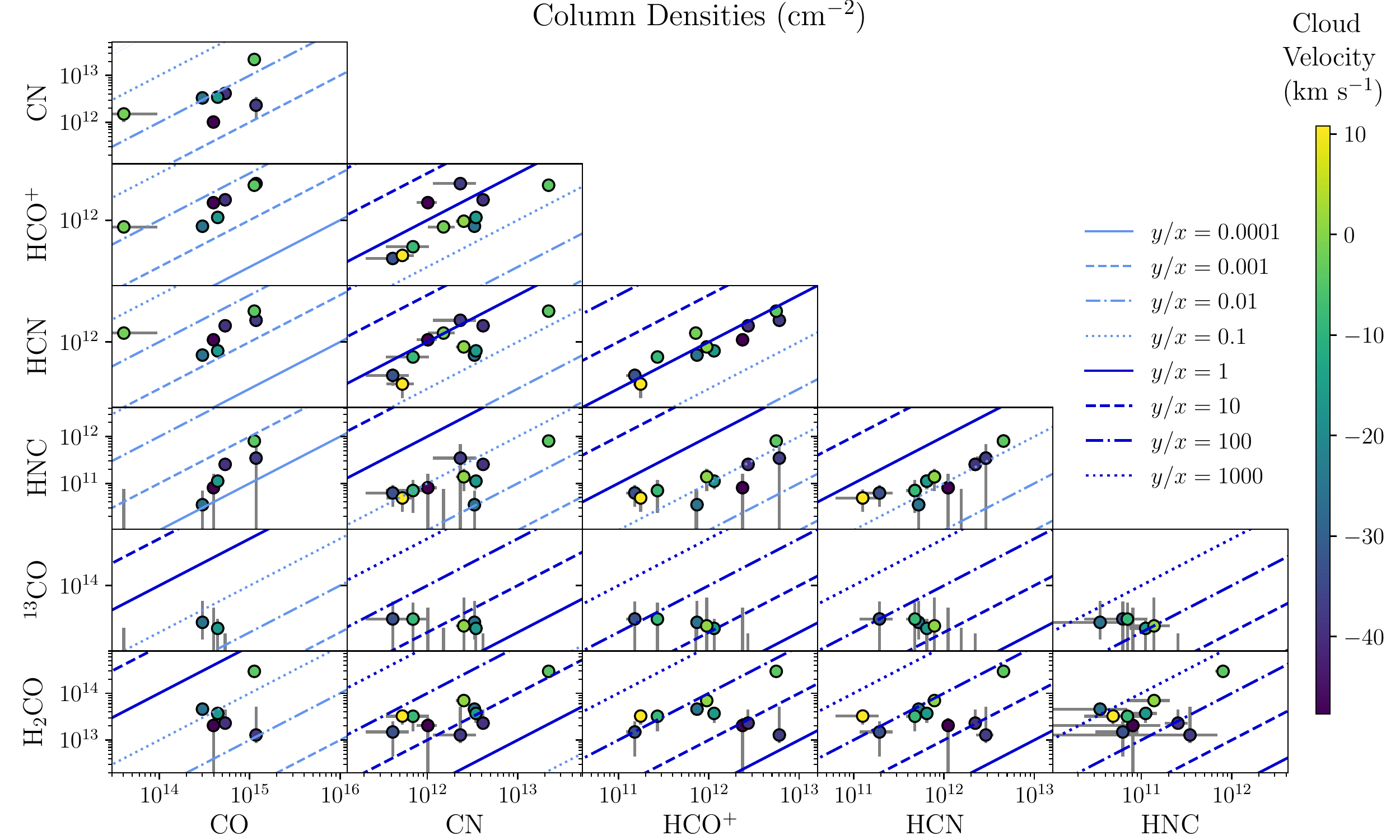}
     \caption{A comparison of the line of sight column densities of CO, CN, HCO$^{+}$, HCN, HNC, $^{13}$CO and H$_{2}$CO. The column densities are calculated from the 12 Gaussian fits applied to the absorption profiles shown in Figs. \ref{fig:MultiSpectra_Plot_diatomic} and \ref{fig:MultiSpectra_Plot_polyatomic}, using Equation \ref{eq:true_column_densities}. The excitation temperature assumed for each absorption region is that estimated in \S\ref{sec:temp_from_HCO} and given in Table \ref{tab:FWHM_v_cen_temp_values}. The colour of each point represents the central velocity of the absorption region relative to the stellar recession velocity of the galaxy, which itself is a good approximation for the velocity of the central supermassive black hole. For CO, \HCO and HCN, which were observed with both the (1-0) and (2-1) rotational lines, the column densities are calculated using the (2-1) line in which the absorption is best resolved.}
    \label{fig:column_densities}
\end{figure*}

The line of sight column densities of the molecular species whose absorption spectra are shown in Figs. \ref{fig:MultiSpectra_Plot_diatomic} and \ref{fig:MultiSpectra_Plot_polyatomic} are given in Tables \ref{tab:mc_fits_tablea} and \ref{tab:mc_fits_tableb}, where the assumed excitation temperature for each of the 12  absorption regions is equal to that calculated as described in \S\ref{sec:temp_from_HCO} and shown in Table \ref{tab:FWHM_v_cen_temp_values}. Using these bespoke excitation temperatures tightens the correlation seen in the column densities compared with when a fixed excitation temperature is assumed for all absorption regions. A corner plot showing how these column densities correlate to one another can be seen in Fig. \ref{fig:column_densities}. Where a molecular species has been observed with multiple rotational lines, e.g. CO(1-0) and CO(2-1), the column densities shown in Fig. \ref{fig:column_densities} are those calculated from the better resolved (2-1) line.

\section{Discussion}
\subsection{A comparison to Milky Way and extragalactic absorption profiles}

\begin{figure*}
	\includegraphics[width=\textwidth]{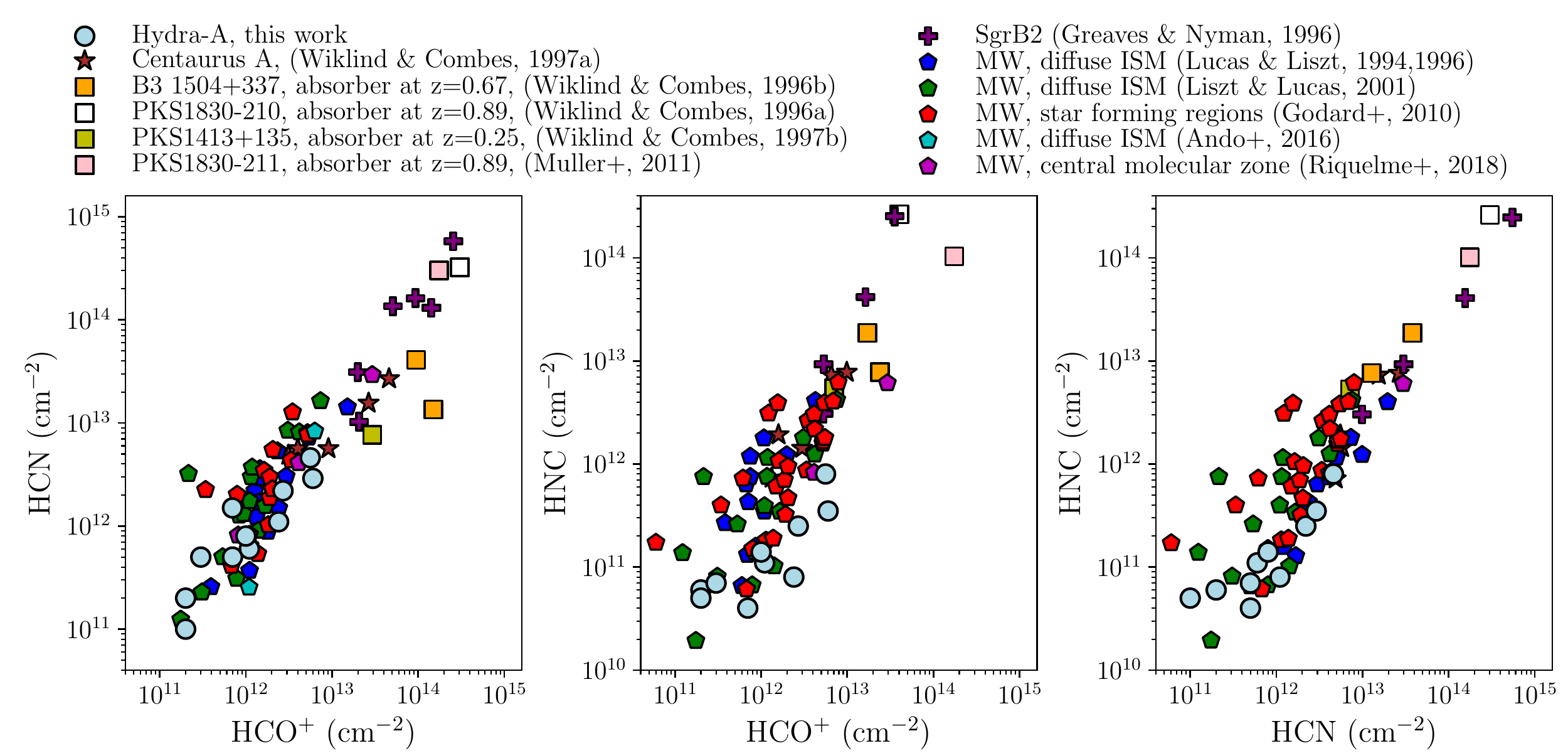}
     \caption{The column densities of HCO$^{+}$, HCN and HNC seen in Hydra-A (circles), Centaurus-A (stars), intervening absorbers i.e. extragalactic sources lying in front of background quasars (squares), Sagittarius B2 (crosses), and the Milky Way (pentagons). The original data are taken from \citet{Wiklind1997a, Wiklind1996a, Wiklind1996b, Wiklind1997b, Muller2011, Greaves1996, Lucas1994, Lucas1996, Liszt2001, Godard2010, Ando2016, Riquelme2017}.}
    \label{fig:ColDensitiesComparison}
\end{figure*}

\begin{figure}
	\includegraphics[width=\columnwidth]{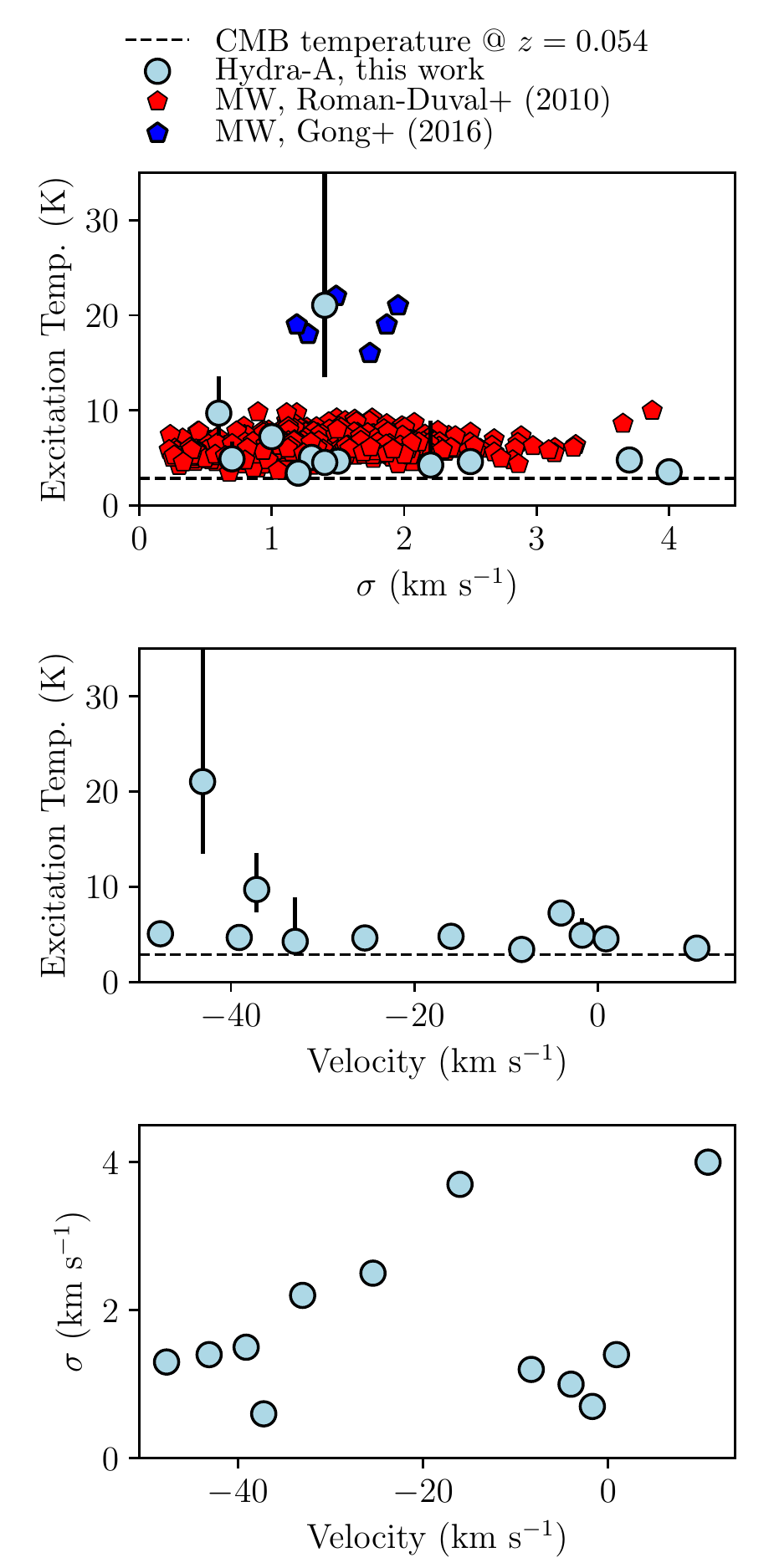}
     \caption{A comparison of the velocity dispersions, $\sigma$, excitation temperatures and line of sight velocities of the absorbing clouds traced by the absorption profiles shown in Figs. \ref{fig:MultiSpectra_Plot_diatomic} and \ref{fig:MultiSpectra_Plot_polyatomic}. The excitation temperatures are those derived from the HCO$^{+}$(1-0) and HCO$^{+}$(2-1) spectra. In the top plot, we show molecular clouds of the Milky Way observed at galactocentric radii of 4 - 8 kpc \citep[red pentagons,][]{RomanDuval2010} and those in star forming regions toward the galactic plane \citep[blue pentagons,][]{Gong2016}.}
    \label{fig:velocity_sigma_temperature_triptych}
\end{figure}

\begin{figure*}
	\includegraphics[width=\textwidth]{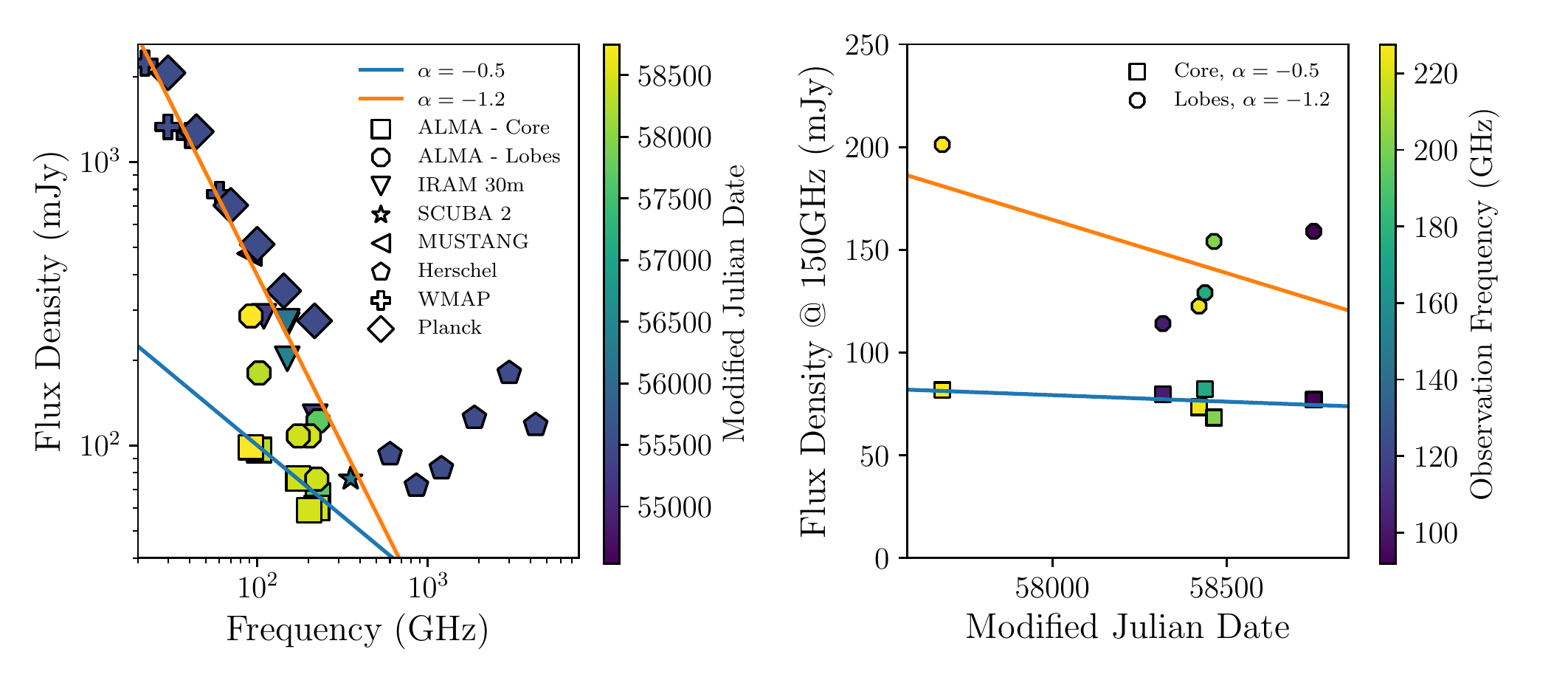}
     \caption{\textbf{Left}: The spectral energy distribution of Hydra-A, produced with data taken since March 2008 using a range of observatories. With the exception of those from ALMA, the observations are of a low angular resolution and consequently include flux from both the radio core and radio lobes. The orange and blue lines show power law fits to the core plus radio lobes and to the resolved core. The increase in emission at $10^{3}$ GHz is from infrared emission due to dust heating. \textbf{Right}: Six ALMA flux density measurements of Hydra-A, all adjusted to give the implied flux density at 150 GHz assuming a power-law spectrum with $\alpha = -0.5$ for the radio core and $\alpha = -1.2$ for the radio lobes. This shows a stable continuum flux density from Hydra-A's core and from its lobes (more scatter is seen in the flux density of the lobes because in some cases, they spread out close to the edge of the observation's the field of view where beam corrections are large).}
    \label{fig:sed}
\end{figure*}

Fig. \ref{fig:ColDensitiesComparison} shows the HCO$^{+}$, HCN and HNC column densities of the molecular clouds in Hydra-A, as well as those found in the Milky Way and other extragalactic sources up to relatively high redshifts of $z=0.89$. The column densities correlate well with those seen in the Milky Way, but are typically lower than in intervening absorber systems by two to three orders of magnitude. A difference this large is not likely to be due to the high quality of the ALMA observations or lower than assumed covering factors; even if the 12 absorption regions were combined, this would still place the column densities at the low end of the scale.

In Fig. \ref{fig:velocity_sigma_temperature_triptych} we show a comparison of the velocity dispersion, excitation temperature and line of sight velocity for each of the 12 absorption regions. Included in the top panel are molecular clouds toward the galactic plane of the Milky Way, as well as those at radii of 4 - 8 kpc. This highlights further similarities between the absorption regions of Hydra-A, which reside in the high pressure environment of a brightest cluster galaxy, and those in the Milky Way. Although we have no strong indication of the distances of Hydra-A's molecular clouds from the centre of the galaxy, this suggests that the locations in which these two sets of clouds reside are fairly interchangeable and that their self-gravitation is significantly more important than the ambient pressures. The properties of the molecular clouds seen in Hydra-A and the Milky way are also both similar to those predicted by accretion simulations \citep[e.g. those of][]{Gaspari2017}.

If the absorbing regions of molecular gas lie on elliptical orbits, their velocities could have apparent shifts relative to the galaxy's systemic velocity of up to a few tens of km s$^{-1}$. In the Keplerian regime, the most blueshifted absorption should lie closest to the galaxy centre. In turn, this could produce a trend between the cloud excitation temperature and velocity as a result of heating from the AGN. Although in Hydra-A the highest excitation temperatures are seen in the most blueshifted clouds, the trend is not especially strong (see the centre panel of Fig. \ref{fig:velocity_sigma_temperature_triptych}). This correlation has been observed within the Milky Way, though it is weak and only visible with the detection of hundreds of molecular clouds \citep{RomanDuval2010}.

The velocity dispersions of most clouds, shown in the lower plot of Fig. \ref{fig:velocity_sigma_temperature_triptych}, are very narrow and lie between 0.5 and 1.4 km s$^{-1}$, indicating that they are due to individual molecular clouds. The outlying absorption regions with higher velocity dispersions are likely small associations of molecular clouds which are not resolved by the observations.

The absorption profiles seen in Hydra-A also bear a strong resemblance to those seen in other systems such as Centaurus-A, and the less well studied brightest cluster galaxy NGC6868 \citep{Israel1990, Rose2019b}. In all three cases there are two deep absorption lines separated by $\sim 50 - 100$ \kms, as well as a more extended absorption complex. Like Hydra-A, Centaurus-A also has a close to edge-on molecular gas disc and an extremely compact core \citep{Israel1990}.

\subsection{Cloud size and mass estimates}
\label{sec:size_mass_estimates}
The similarities between the clumpy interstellar medium of the Milky Way and that which we see along our line of sight to the core of Hydra-A allow us derive estimates of the size and mass of the molecular clouds observed. Within the Milky Way, the velocity dispersion, $\sigma$, and diameter, D, of molecular clouds are related by:
\begin{equation}
     \left(\frac{\sigma}{\textnormal{km s}^{-1}}\right) = \left(\frac{D}{\textnormal{pc}}\right)^{0.5}.
\end{equation}{}
This relation was first shown by \citet{Larson1981} and its approximate form has been supported by many more recent works \citep[e.g.][]{Solomon1987, Vazquez-Semadeni2007, McKee2007, Ballesteros-Paredes2011}. Hydra-A is a brightest cluster galaxy and so its thermal pressure is many times higher than that of the Milky Way. If the molecular gas behaves in a reactive way to this different environment, then the relation may be less applicable. However, as Fig. \ref{fig:ColDensitiesComparison} shows, the clouds' environment does not result in significantly higher line of sight column densities and so the relation should still hold true.
The implied sizes of the 12 absorption regions detected in Hydra-A are given in Table \ref{tab:FWHM_v_cen_temp_values}.

By further assuming the absorption regions are in virial equilibrium, their total masses, $M_{\textnormal{tot}}$, can be estimated using the virial theorem:

\begin{equation}
    M_{\textnormal{tot}} = \frac{D \sigma^{2}}{2G},
\label{eq:virial_theorem}
\end{equation}
where $D$ is the cloud diameter and $G$ is the gravitational constant. The total masses of the 12 absorption regions are given in Table \ref{tab:FWHM_v_cen_temp_values}.

\subsection{An estimate of the continuum source's size}

The above estimate of the total cloud mass can be used in conjunction with the line of sight column density of molecular hydrogen derived from X-ray observations to estimate the size of the continuum source against which absorption is seen. This will only provide a rough estimate due to the uncertainties in the cloud masses and the likely difference in the size of the continuum source between the frequencies of the X-ray and radio observations.

The total mass inferred from the molecular absorption seen in Hydra-A is $4\times 10 ^{4} M_{\odot}$ and \citet{Russell2013} find a line of sight column density of $N_{H} = 3.5 \times 10^{22} \textnormal{cm}^{-2}$ from X-ray observations. These values imply that Hydra-A's central continuum source has an apparent diameter of 7 pc, assuming it appears circular along the line of sight.

The above value is likely an overestimate because most of the mass we estimate in \S\ref{sec:size_mass_estimates} comes from the broadest absorption regions. These are unlikely to be individual molecular clouds in virial equilibrium, but rather collections of unresolved molecular clouds. To make some correction for this, we use the very simple assumption that the widest clouds, G7 and G12, are each in fact the combination of two molecular clouds, with a velocity dispersion half of the original value. This reduces their estimated diameters by a factor of four, and their masses by a factor of sixteen. There are now twice as many clouds, so the overall mass of these absorption features is reduced by a factor of eight. This new mass results in an estimated continuum diameter of 4 pc. VLBA observations by \citet{Taylor1996} at the lower frequency 1.35 GHz show hints of structure on similar scales.

\subsection{Continuum variability}

Hydra-A has been attentively studied at a wide range of frequencies over several decades. The left panel of Fig. \ref{fig:sed} shows the galaxy's spectral energy distribution and the right shows the continuum variability of its core and radio lobes, as seen with the ALMA observations presented in this paper.

The left panel of Fig. \ref{fig:sed} shows that no significant change in the flux density of the core, against which the absorption is detected, has taken place over the two year time range in which the observations were taken. The flux density of the lobes is expected to be constant and the significant scatter present is a result of the limited angular resolution of the observations, with the lobes often spreading out close to the edge of the field of view where beam corrections are large.

\begin{figure*}
	\includegraphics[width=\textwidth]{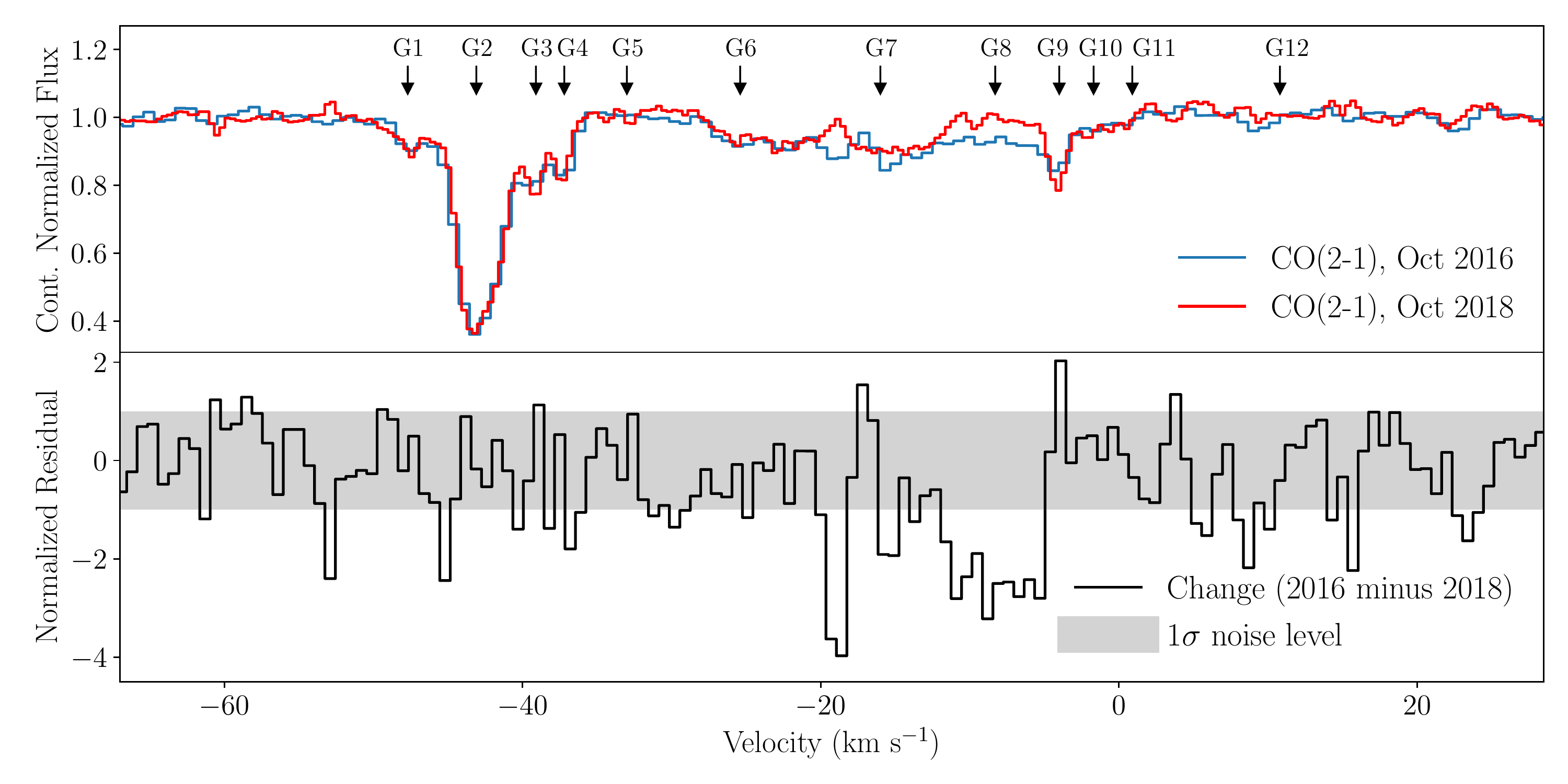}
     \caption{\textbf{Top:} The overlaid absorption profiles of two CO(2-1) spectra taken in October 2016 and October 2018. The two spectra are extracted from a region with a size equal to the synthesized beam's FWHM centred on Hydra-A's bright and compact radio core. The absorption is largely consistent given the noise levels, though a small difference appears between $-20$ and $-5$ km s$^{-1}$. The G1-12 markers indicate the central velocities of each component of the 12-Gaussian fit made to the spectra in Figs. \ref{fig:MultiSpectra_Plot_diatomic} and \ref{fig:MultiSpectra_Plot_polyatomic}. \textbf{Bottom:} The change seen in the absorption between the two observations, with the grey band indicating the 1$\sigma$ noise level of the residual. The variability of the spectrum over the velocity range in which absorption is seen (approximately -50 to 10 km s$^{-1}$) has a combined significance of $3.4 \sigma$, as calculated from a $\chi^{2}$ distribution test.}
    \label{fig:Overlapping_CO_21}
\end{figure*}

\subsection{Absorption variability}

CO(2-1) absorption was first detected in Hydra-A in October 2016 \citep{Rose2019a}, and a repeat observation was carried out in October 2018 as part of the main survey presented in this paper. In Fig. \ref{fig:Overlapping_CO_21} we show the absorption profiles seen against the galaxy's bright radio core for these two observations. Across the majority of the absorption profile, the two spectra are consistent within the noise levels and there is little hint of variability. However, between $-20$ and $-5$ km s$^{-1}$ a decrease in the absorption strength appears to have taken place over the intervening two years. This velocity range does not correspond well to any component of the multi-Gaussian best fit and so is unlikely to be due to a change in the absorption of any individual molecular cloud. The variability of the spectrum over the velocity range in which absorption is seen (approximately -50 to 10 km s$^{-1}$) has a combined significance of $3.4 \sigma$ (calculated from a $\chi^{2}$ distribution test)\footnote{A shift of $\sim 0.1$ mJy has been applied to the spectrum extracted from the 2018 observation. This error appears as a result of the subtraction of the emission from the two spectra, where the degree to which the emission compensates for the absorption cannot be known precisely \citep[see][fig. 6 for a plot showing the CO(2-1) emission from the core]{Rose2019a}}.

If this apparent change in the absorption profile is real, then it will almost certainly be due to either a change in the continuum source against which the absorption is observed, or due to a movement of the molecular clouds responsible for the absorption. Below we discuss these explanations, both of which are likely to be true to a greater or lesser extent.

The unresolved continuum source we observe the absorption against may be made up of several components, each covered to varying degrees by different gas clouds. This would produce absorption along multiple lines of sight, which then combines to make the single absorption profile we observe against the unresolved continuum source. This is consistent with 1.35 GHz VLBA observations by \citet{Taylor1996}, which show the continuum source at high angular resolution. Spatially resolved H\thinspace\small{I}\normalsize\space absorption is seen, most likely caused by different lines of sight toward the radio core and the knots of the galaxy's jets. If this is the case, we would detect no absorption which appears to be optically thick, and the reduced strength of the absorption may be due to a decrease in flux from one particular component of the continuum source. This would result in weaker absorption from that line of sight, but leave the remaining absorption unaffected. Although there is no significant change in the continuum's strength over this time period (see Fig. \ref{fig:sed}), the dimming required to produce the small decrease in absorption could well be within the noise levels of the measurements of the continuum flux density. Even if the total continuum emission is not varying, relativistic and transverse motions in the hot spots of the continuum source could change the background illumination of the absorbing clouds.

An angular precession of the continuum source could also result in a change in the background illumination of the molecular gas. \citet{Nawaz2016} found a precession period of $\sim$ 1 Myr in Hydra-A from hydrodynamical simulations of its jet-intracluster medium interactions. In a two year time frame this translates to an angular precession of 2.6", which sweeps over a transverse distance of 0.01 pc at a radius of 1 kpc from the continuum source, or 0.1 pc at a radius of 10 kpc. Given the typical size of the molecular clouds of around 1 pc (see Table \ref{tab:FWHM_v_cen_temp_values}) and that they likely have a fractal substructure, the latter seems plausible. However, with a precessing continuum source the level of variability would increase as the distance of the clouds from the nucleus increases. To observe a significant level of variability between $-20$ and $-5$ km s$^{-1}$, but little hint of it anywhere else requires there to be a group of clouds close to the continuum source where the angular change encompasses a negligible linear scale relative to the cloud size, and one group at very large distances, with no clouds in between. Further, the edge on disc of Hydra-A has a radius of around 2.5 kpc, so at 10 kpc the column density of cold molecular gas present is likely low compared with smaller radii.

A variation in the absorption could also be produced by transverse movement of the molecular gas responsible for the absorption between $-20$ and $-5$ km s$^{-1}$. However, even a relatively small molecular cloud with a diameter of 0.1 pc and a large transverse velocity of 500 \kms will take $\sim$ 200 years to fully cross the line of sight, assuming a point-like continuum source. Particularly small and dense, inhomogeneous, or fast moving molecular clouds would therefore be required for this effect to be seen within a two year time frame. The transverse velocities of molecular clouds are orders of magnitude less than the relativistic and potentially superluminal motions of the knots in the jets at the core of the continuum source, so this can ruled out with a fair degree of confidence. 

Alterations in the cloud chemistry could also result in variability. Although this can occur on cosmologically short timescales on the order of $10^{5}$ years \citep{Haranda2019}, this is much greater than the two year interval over which we detect variability. Only a change induced by significant alterations to the local cosmic ray field, for example, by a nearby supernova could occur quickly enough. Although physically feasible, this `by chance' explanation in unlikely given that absorption variability has also been seen in several similar intervening absorber systems \citep[e.g.][]{Wiklind1997b, Muller2011}.


When considering the above explanations it should be noted that clear absorption is still seen within this velocity range in CN(2-1), HCO$^{+}$(2-1) and HCN$^{+}$(2-1), though this may well have been stronger still if all of the observations had been taken in October 2016 rather than October 2018. Further observations of these lines, where this absorption is strongest and any changes would be more evident, would therefore track any variability in more detail and reveal its cause. 

\section{Conclusions}

We present ALMA observations of CO, $^{13}$CO, CN, SiO, HCO$^{+}$, HCN, HNC and H$_{2}$CO molecular absorption lines seen against the bright radio core of Hydra-A. Their narrow velocity dispersions (typically $\sim 1 \textnormal{km s}^{-1}$) are similar to those seen molecular cloud complexes of the Milky Way and indicate that the observations are tracing individual clouds of cold molecular gas. The molecular gas clouds typically have excitation temperatures of 5 - 10 K, diameters of 1 - 10 pc and masses of a few tens to a few thousands of $M_{\odot}$. 

The precise origins and locations of the absorbing molecular clouds within Hydra-A are difficult to constrain, though they are most likely to be within the inner few kpc of the disc where the column densities of molecular gas are highest. The observations are evidence of a clumpy interstellar medium, consistent with galaxy-wide fuelling and feedback cycles predicted by e.g. \citet{Pizzolato2005, PetersonFabian2006, McNamara2016, Gaspari2018}.

Future surveys targeting molecular absorption in brightest cluster galaxies are likely to be most successful when searching for \HCO and HCN lines, which we find to be stronger, more ubiquitous and more consistent tracers than CO. A further advantage of these molecules over CO is a lack of significant emission, which can counteract any absorption and make the true optical depths unclear. HCO$^{+}$ is particularly useful because it lacks any hyperfine structure.

We have compared the line of sight column densities, velocity dispersions and excitation temperatures of the molecular clouds seen in Hydra-A to those of the Milky Way. The two populations are largely indistinguishable, implying that the high pressure environment of a brightest cluster galaxy has negligible effect on the molecular clouds when compared with their self-gravitation. 

The line of sight absorption seen against Hydra-A's bright radio core has shown variation at 3.4 $\sigma$ significance between ALMA Cycle 4 and 6 observations. These observations are separated by two years, so if this variability is genuine it is occurring on galactically short timescales. The first of two likely explanations for the variability is a multi-component continuum source, one component of which has decreased in brightness or has seen relativistic movement in a hot spot, in turn giving decreased absorption along one particular line of sight. A second possible but less likely explanation is that one of the many absorbing clouds, or groups of absorbing clouds, has significant transverse motion such that it no longer covers the continuum source in the same way.

\section*{Acknowledgements}
We thank the referee for their time and comments, which have helped us to improve the paper. 
We are grateful to Rick Perley for providing the VLA image used in Fig. \ref{fig:Hydra-A_HST_cont_image}.

T.R. is supported by the Science and Technology Facilities Council (STFC) through grant ST/R504725/1.

A.C.E. acknowledges support from STFC grant ST/P00541/1.

M.G. is supported by the Lyman Spitzer Jr. Fellowship
(Princeton University) and by NASA Chandra GO8-19104X/GO9-20114X and HST GO-
15890.020-A grants.

This paper makes use of the following ALMA data: ADS/JAO.ALMA\#2016.1.01214.S, ADS/JAO.ALMA\#2017.1.00629.S and ADS/JAO.ALMA\#2018.1.01471.S. ALMA is a partnership of ESO (representing its member states), NSF (USA) and NINS (Japan), together with NRC (Canada), MOST and ASIAA (Taiwan), and KASI (Republic of Korea), in cooperation with the Republic of Chile. The Joint ALMA Observatory is operated by ESO, AUI/NRAO and NAOJ.



\appendix 

\section{Hyperfine structure of CN(2-1) and HCN(2-1)}
\label{sec:CN_HCN_hfs}

\begin{figure*}
	\includegraphics[width=\textwidth]{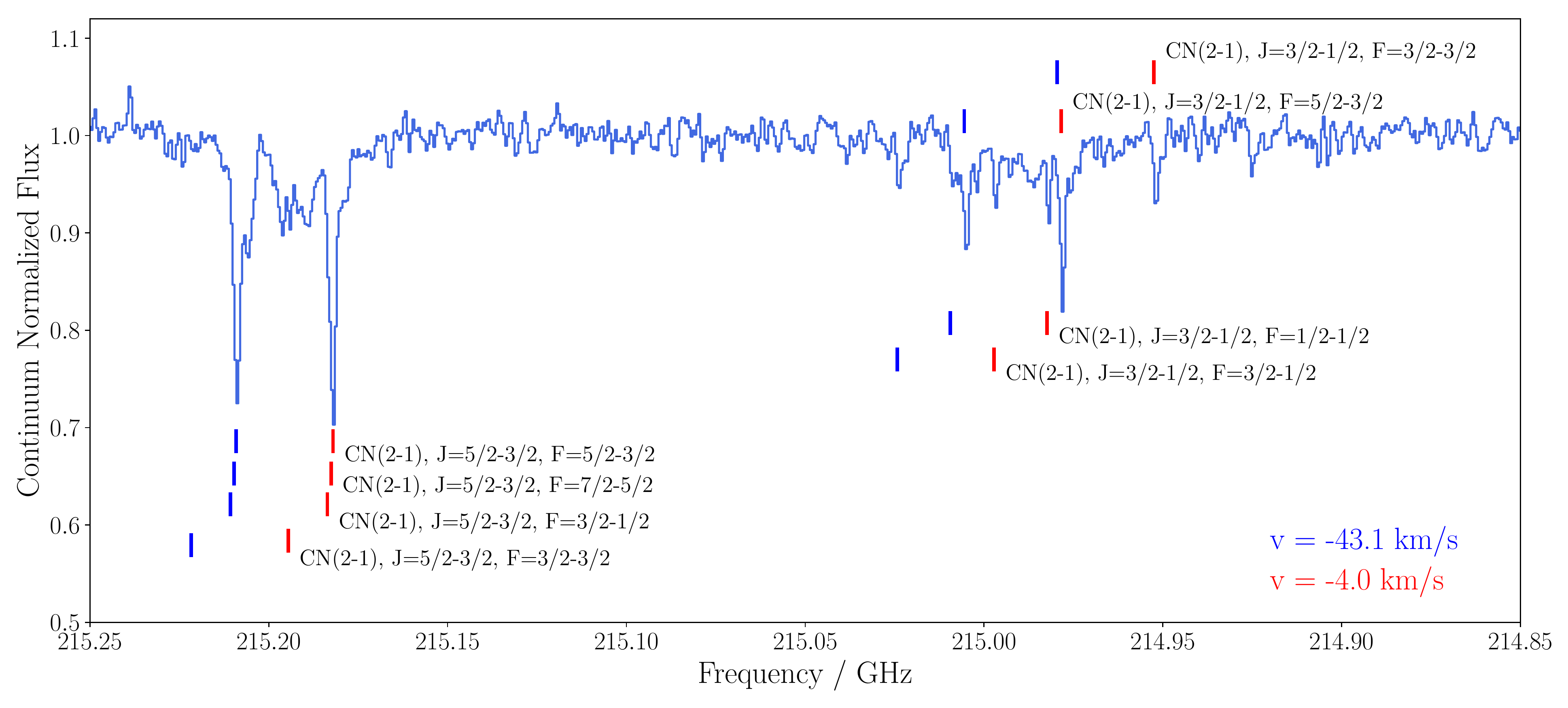}
     \caption{The spectrum of CN(2-1) seen against the line of sight to Hydra-A's bright radio core, which contains absorption from several of the molecule's hyperfine structure lines. The intensity and rest frequencies of these hyperfine structure lines are given in Table \ref{tab:CN_rotational_lines}. Markers on the plot indicate where absorption would be expected from the dominant hyperfine structure lines due to the two strongest points of absorption at $-43.1$ and $-4.0$ km s$^{-1}$. When analysing the CN(2-1) absorption in the main body of the paper, we focus on the combination of hyperfine structure lines seen between approximately $215.15 - 215.25$ GHz and disregard the rest.}
    \label{fig:CN_Potential_Lines_Plot}
\end{figure*}

The spectra of the CN(2-1) and HCN(2-1) observations contain absorption features of several hyperfine structure lines. The frequencies of these lines and their relative strengths are given in Tables \ref{tab:CN_rotational_lines} and \ref{tab:HCN_rotational_lines}. 

In Fig. \ref{fig:CN_Potential_Lines_Plot} we show the full spectrum of CN(2-1), as well as markers which indicate where absorption would be expected due to the hyperfine structure lines for the strong absorption features at $-43.1$ and $-4.0$ \kms (see Figs. \ref{fig:MultiSpectra_Plot_diatomic} and \ref{fig:MultiSpectra_Plot_polyatomic}). In the main body of the paper we focus on the strongest set of absorption features which can be seen at approximately $215.15 - 215.25$ GHz and disregard the rest (note that in Fig. \ref{fig:CN_Potential_Lines_Plot} the frequency axis is reversed such that the direction of left to right implies increasing velocity for each of the lines, in keeping with the paper's other plots). This set of absorption features is produced by the combination of four hyperfine structure lines. However, three of these lines are very close to each other in frequency and the fourth is of negligible strength (5 per cent of the other three lines combined) and so is ignored for simplicity.

Although the three remaining hyperfine structure lines are very close to each other in frequency, the difference between them does have a tangible effect on the apparent velocity dispersion, $\sigma$, of the absorption features. We therefore make a slight modification to the rigid Gaussian fitting process used for the other spectra which have no or negligible hyperfine structure. When fitting to all of the other spectra we use a 12-Gaussian line, where each Gaussian has a fixed $v_{\textnormal{cen}}$ and $\sigma$, but varying amplitude. To estimate the effect of the CN molecule's hyperfine structure, we simulate each of the 12 Gaussian features as they would appear due to the three overlapping hyperfine structure lines. This produces what appears to be a single, stronger Gaussian absorption line with a slightly wider $\sigma$ than the individual lines. This slightly wider $\sigma$ is due the separation of the hyperfine structure lines. When a single Gaussian is fitted to each of these simulated absorption features, we find the increase in the $\sigma$ caused by the hyperfine structure (typically $0.1 -0.3$ km s$^{-1}$). When making the 12-Gaussian fit to the CN(2-1) spectrum shown in Fig. 3, we therefore increase the $\sigma$ of each component Gaussian line accordingly. This process has also been tested for HCN(2-1), which contains similar hyperfine structure. However, we find and unmodified fit works best, most likely because for HCN(2-1), around 70 percent of the flux is contained within two hyperfine lines just 0.0001 GHz apart.

\begin{table*}
	\centering
	\begin{tabular}{ccc} 

		\hline
		Rest frequency (GHz) & CN Transition & Relative intensity \\
		\hline
226.28741850 & J=3/2-3/2, F=1/2-1/2 &	0.0060 \\
226.29894270 & J=3/2-3/2, F=1/2-3/2 &	0.0048 \\
226.30303720 & J=3/2-3/2, F=3/2-1/2 &	0.0049 \\
226.31454000 & J=3/2-3/2, F=3/2-3/2 &	0.0116 \\
226.33249860 & J=3/2-3/2, F=3/2-5/2 &	0.0053 \\
226.34192980 & J=3/2-3/2, F=5/2-3/2 &	0.0055 \\
226.35987100 & J=3/2-3/2, F=5/2-5/2 &	0.0282 \\
226.61657140 & J=3/2-1/2, F=1/2-3/2 &	0.0063 \\
226.63219010 & J=3/2-1/2, F=3/2-3/2 &	0.0498 \\
226.65955840 & J=3/2-1/2, F=5/2-3/2 &	0.1660 \\
226.66369280 & J=3/2-1/2, F=1/2-1/2 &	0.0495 \\
226.67931140 & J=3/2-1/2, F=3/2-1/2 &	0.0616 \\
226.87419080 & J=5/2-3/2, F=5/2-3/2 &	0.1685 \\
226.87478130 & J=5/2-3/2, F=7/2-5/2 &	0.2669 \\
226.87589600 & J=5/2-3/2, F=3/2-1/2 &	0.1002 \\
226.88742020 & J=5/2-3/2, F=3/2-3/2 &	0.0319 \\
226.89212800 & J=5/2-3/2, F=5/2-5/2 &	0.0317 \\
226.90535740 & J=5/2-3/2, F=3/2-5/2 &	0.0013 \\
		\hline 
		
	\end{tabular}
    \caption{Hyperfine structure lines of CN(2-1) \citep{Splatalogue_CDMS}.}
    \label{tab:CN_rotational_lines}
\end{table*}

\begin{table*}
	\centering
	\begin{tabular}{ccc} 

		\hline
		Rest frequency (GHz) & HCN Transition & Relative intensity \\
		\hline
         88.63393600 & J=1-0, F=0-1 & 0.1111 \\
         88.63041600 & J=1-0, F=1-1 & 0.3333 \\
         88.63184700 & J=1-0, F=2-1 & 0.5556 \\
		\hline 
		177.25967700 & J=2-1,F=2-2 & 0.0833\\
		177.25992300 & J=2-1,F=1-0 & 0.1111\\
		177.26111000 & J=2-1,F=2-1 & 0.2500 \\
		177.26122300 & J=2-1,F=3-2 & 0.4667\\
		177.26201220 & J=2-1,F=1-2 & 0.0056\\
		177.26344500 & J=2-1,F=1-1 & 0.0833\\
		\hline
		
	\end{tabular}
    \caption{Hyperfine structure lines of HCN(1-0) and HCN(2-1) \citep{Splatalogue_CDMS}.}
    \label{tab:HCN_rotational_lines}
\end{table*}

\bsp	
\label{lastpage}

\begin{thebibliography}{}
\makeatletter
\relax
\def\mn@urlcharsother{\let\do\@makeother \do\$\do\&\do\#\do\^\do\_\do\%\do\~}
\def\mn@doi{\begingroup\mn@urlcharsother \@ifnextchar [ {\mn@doi@}
  {\mn@doi@[]}}
\def\mn@doi@[#1]#2{\def\@tempa{#1}\ifx\@tempa\@empty \href
  {http://dx.doi.org/#2} {doi:#2}\else \href {http://dx.doi.org/#2} {#1}\fi
  \endgroup}
\def\mn@eprint#1#2{\mn@eprint@#1:#2::\@nil}
\def\mn@eprint@arXiv#1{\href {http://arxiv.org/abs/#1} {{\tt arXiv:#1}}}
\def\mn@eprint@dblp#1{\href {http://dblp.uni-trier.de/rec/bibtex/#1.xml}
  {dblp:#1}}
\def\mn@eprint@#1:#2:#3:#4\@nil{\def\@tempa {#1}\def\@tempb {#2}\def\@tempc
  {#3}\ifx \@tempc \@empty \let \@tempc \@tempb \let \@tempb \@tempa \fi \ifx
  \@tempb \@empty \def\@tempb {arXiv}\fi \@ifundefined
  {mn@eprint@\@tempb}{\@tempb:\@tempc}{\expandafter \expandafter \csname
  mn@eprint@\@tempb\endcsname \expandafter{\@tempc}}}

\bibitem[\protect\citeauthoryear{{Ando}, {Kohno}, {Tamura}, {Izumi}, {Umehata}
  \& {Nagai}}{{Ando} et~al.}{2016}]{Ando2016}
{Ando} R.,  {Kohno} K.,  {Tamura} Y.,  {Izumi} T.,  {Umehata} H.,   {Nagai} H.,
   2016, \mn@doi [\pasj] {10.1093/pasj/psv110}, \href
  {https://ui.adsabs.harvard.edu/abs/2016PASJ...68....6A} {68, 6}

\bibitem[\protect\citeauthoryear{Ballesteros-Paredes, Hartmann,
  Vazquez-Semadeni, Heitsch  \& Zamora-Aviles}{Ballesteros-Paredes
  et~al.}{2011}]{Ballesteros-Paredes2011}
Ballesteros-Paredes J.,  Hartmann L.~W.,  Vazquez-Semadeni E.,  Heitsch F.,
  Zamora-Aviles M.~A.,  2011, \mn@doi [\mnras]
  {10.1111/j.1365-2966.2010.17657.x}, 411, 65

\bibitem[\protect\citeauthoryear{Boger \& Sternberg}{Boger \&
  Sternberg}{2005}]{Boger2005}
Boger G.~I.,  Sternberg A.,  2005, \mn@doi [\apj] {10.1086/432864}, 632, 302

\bibitem[\protect\citeauthoryear{{Bolatto}, {Leroy}, {Israel}  \&
  {Jackson}}{{Bolatto} et~al.}{2003}]{Bolatto2003}
{Bolatto} A.~D.,  {Leroy} A.,  {Israel} F.~P.,   {Jackson} J.~M.,  2003,
  \mn@doi [\apj] {10.1086/377230}, \href
  {http://adsabs.harvard.edu/abs/2003ApJ...595..167B} {595, 167}

\bibitem[\protect\citeauthoryear{{Brown} \& {Wilson}}{{Brown} \&
  {Wilson}}{2019}]{Brown2019}
{Brown} T.,  {Wilson} C.,  2019, arXiv e-prints, \href
  {https://ui.adsabs.harvard.edu/abs/2019arXiv190506950B} {p. arXiv:1905.06950}

\bibitem[\protect\citeauthoryear{{Combes}, {Gupta, N.}, {Jozsa, G. I. G.}  \&
  {Momjian, E.}}{{Combes} et~al.}{2019}]{Combes2019}
{Combes} {Gupta, N.} {Jozsa, G. I. G.}  {Momjian, E.} 2019, \mn@doi [A\&A]
  {10.1051/0004-6361/201935057}, 623, A133

\bibitem[\protect\citeauthoryear{David et~al.,}{David et~al.}{2014}]{David2014}
David L.~P.,  et~al., 2014, \apj, 792, 94

\bibitem[\protect\citeauthoryear{{Davis}}{{Davis}}{2014}]{Davis2014}
{Davis} T.~A.,  2014, \mn@doi [\mnras] {10.1093/mnras/stu1850}, \href
  {https://ui.adsabs.harvard.edu/abs/2014MNRAS.445.2378D} {445, 2378}

\bibitem[\protect\citeauthoryear{{Donahue}, {de Messi{\`e}res}, {O'Connell},
  {Voit}, {Hoffer}, {McNamara}  \& {Nulsen}}{{Donahue}
  et~al.}{2011}]{Donahue2011}
{Donahue} M.,  {de Messi{\`e}res} G.~E.,  {O'Connell} R.~W.,  {Voit} G.~M.,
  {Hoffer} A.,  {McNamara} B.~R.,   {Nulsen} P.~E.~J.,  2011, \mn@doi [\apj]
  {10.1088/0004-637X/732/1/40}, \href
  {http://adsabs.harvard.edu/abs/2011ApJ...732...40D} {732, 40}

\bibitem[\protect\citeauthoryear{{Downes}, {Wilson}, {Bieging}  \&
  {Wink}}{{Downes} et~al.}{1980}]{Downes1980}
{Downes} D.,  {Wilson} T.~L.,  {Bieging} J.,   {Wink} J.,  1980, \aaps, \href
  {https://ui.adsabs.harvard.edu/abs/1980A%26AS...40..379D} {40, 379}

\bibitem[\protect\citeauthoryear{{Dwarakanath}, {Owen}  \& {van
  Gorkom}}{{Dwarakanath} et~al.}{1995}]{Dwarakanath1995}
{Dwarakanath} K.~S.,  {Owen} F.~N.,   {van Gorkom} J.~H.,  1995, \mn@doi
  [\apjl] {10.1086/187801}, \href
  {http://adsabs.harvard.edu/abs/1995ApJ...442L...1D} {442, L1}

\bibitem[\protect\citeauthoryear{{Edge}, {Shakeshaft}, {McAdam}, {Baldwin}  \&
  {Archer}}{{Edge} et~al.}{1959}]{Edge1959}
{Edge} D.~O.,  {Shakeshaft} J.~R.,  {McAdam} W.~B.,  {Baldwin} J.~E.,
  {Archer} S.,  1959, \memras, \href
  {http://adsabs.harvard.edu/abs/1959MmRAS..68...37E} {68, 37}

\bibitem[\protect\citeauthoryear{{Edge}, {Wilman}, {Johnstone}, {Crawford},
  {Fabian}  \& {Allen}}{{Edge} et~al.}{2002}]{Edge2002}
{Edge} A.~C.,  {Wilman} R.~J.,  {Johnstone} R.~M.,  {Crawford} C.~S.,  {Fabian}
  A.~C.,   {Allen} S.~W.,  2002, \mn@doi [\mnras]
  {10.1046/j.1365-8711.2002.05790.x}, \href
  {http://adsabs.harvard.edu/abs/2002MNRAS.337...49E} {337, 49}

\bibitem[\protect\citeauthoryear{{Garc{\'{\i}}a-Burillo}
  et~al.,}{{Garc{\'{\i}}a-Burillo} et~al.}{2014}]{GarciaBurillo2014}
{Garc{\'{\i}}a-Burillo} S.,  et~al., 2014, \mn@doi [\aap]
  {10.1051/0004-6361/201423843}, \href
  {http://adsabs.harvard.edu/abs/2014A%26A...567A.125G} {567, A125}

\bibitem[\protect\citeauthoryear{{Gaspari}, {Temi}  \& {Brighenti}}{{Gaspari}
  et~al.}{2017}]{Gaspari2017}
{Gaspari} M.,  {Temi} P.,   {Brighenti} F.,  2017, \mn@doi [\mnras]
  {10.1093/mnras/stw3108}, \href
  {http://adsabs.harvard.edu/abs/2017MNRAS.466..677G} {466, 677}

\bibitem[\protect\citeauthoryear{Gaspari et~al.}{Gaspari
  et~al.}{2018}]{Gaspari2018}
Gaspari M.,  et~al., 2018, \mn@doi [Astrophys. J.] {10.3847/1538-4357/aaaa1b},
  854, 167

\bibitem[\protect\citeauthoryear{{Gaspari}, {Tombesi}  \& {Cappi}}{{Gaspari}
  et~al.}{2020}]{Gaspari2020}
{Gaspari} M.,  {Tombesi} F.,   {Cappi} M.,  2020, \mn@doi [Nature Astronomy]
  {10.1038/s41550-019-0970-1}, \href
  {https://ui.adsabs.harvard.edu/abs/2020NatAs...4...10G} {4, 10}

\bibitem[\protect\citeauthoryear{{Gerin}, {Liszt}, {Neufeld}, {Godard},
  {Sonnentrucker}, {Pety}  \& {Roueff}}{{Gerin} et~al.}{2019}]{Gerin2019}
{Gerin} M.,  {Liszt} H.,  {Neufeld} D.,  {Godard} B.,  {Sonnentrucker} P.,
  {Pety} J.,   {Roueff} E.,  2019, \mn@doi [\aap]
  {10.1051/0004-6361/201833661}, \href
  {https://ui.adsabs.harvard.edu/abs/2019A&A...622A..26G} {622, A26}

\bibitem[\protect\citeauthoryear{{Ginard} et~al.,}{{Ginard}
  et~al.}{2012}]{Ginard2012}
{Ginard} D.,  et~al., 2012, \mn@doi [\aap] {10.1051/0004-6361/201118347}, \href
  {https://ui.adsabs.harvard.edu/abs/2012A&A...543A..27G} {543, A27}

\bibitem[\protect\citeauthoryear{{Glenn} \& {Hunter}}{{Glenn} \&
  {Hunter}}{2001}]{Glenn2001}
{Glenn} J.,  {Hunter} T.~R.,  2001, \mn@doi [\apjs] {10.1086/321791}, \href
  {https://ui.adsabs.harvard.edu/abs/2001ApJS..135..177G} {135, 177}

\bibitem[\protect\citeauthoryear{{Godard}, {Falgarone}, {Gerin}, {Hily-Blant}
  \& {de Luca}}{{Godard} et~al.}{2010}]{Godard2010}
{Godard} B.,  {Falgarone} E.,  {Gerin} M.,  {Hily-Blant} P.,   {de Luca} M.,
  2010, \mn@doi [\aap] {10.1051/0004-6361/201014283}, \href
  {http://adsabs.harvard.edu/abs/2010A%26A...520A..20G} {520, A20}

\bibitem[\protect\citeauthoryear{{Gong} et~al.,}{{Gong}
  et~al.}{2016}]{Gong2016}
{Gong} Y.,  et~al., 2016, \mn@doi [\aap] {10.1051/0004-6361/201527334}, \href
  {https://ui.adsabs.harvard.edu/abs/2016A&A...588A.104G} {588, A104}

\bibitem[\protect\citeauthoryear{{Greaves} \& {Nyman}}{{Greaves} \&
  {Nyman}}{1996}]{Greaves1996}
{Greaves} J.~S.,  {Nyman} L.~A.,  1996, \aap, \href
  {https://ui.adsabs.harvard.edu/abs/1996A&A...305..950G} {305, 950}

\bibitem[\protect\citeauthoryear{{Hacar}, {Bosman}  \& {van Dishoeck}}{{Hacar}
  et~al.}{2019}]{Hacar2019}
{Hacar} A.,  {Bosman} A.~D.,   {van Dishoeck} E.~F.,  2019, arXiv e-prints,
  \href {https://ui.adsabs.harvard.edu/abs/2019arXiv191013754H} {p.
  arXiv:1910.13754}

\bibitem[\protect\citeauthoryear{{Hamer} et~al.,}{{Hamer}
  et~al.}{2014}]{Hamer2014}
{Hamer} S.~L.,  et~al., 2014, \mn@doi [\mnras] {10.1093/mnras/stt1949}, \href
  {https://ui.adsabs.harvard.edu/abs/2014MNRAS.437..862H} {437, 862}

\bibitem[\protect\citeauthoryear{{Hansen}, {Jorgensen}  \&
  {Norgaard-Nielsen}}{{Hansen} et~al.}{1995}]{Hansen1995}
{Hansen} L.,  {Jorgensen} H.~E.,   {Norgaard-Nielsen} H.~U.,  1995, \aap, \href
  {http://adsabs.harvard.edu/abs/1995A%26A...297...13H} {297, 13}

\bibitem[\protect\citeauthoryear{{Harada}, {Nishimura}, {Watanabe}, {Yamamoto},
  {Aikawa}, {Sakai}  \& {Shimonishi}}{{Harada} et~al.}{2019}]{Haranda2019}
{Harada} N.,  {Nishimura} Y.,  {Watanabe} Y.,  {Yamamoto} S.,  {Aikawa} Y.,
  {Sakai} N.,   {Shimonishi} T.,  2019, \mn@doi [\apj]
  {10.3847/1538-4357/aaf72a}, \href
  {https://ui.adsabs.harvard.edu/abs/2019ApJ...871..238H} {871, 238}

\bibitem[\protect\citeauthoryear{{Henkel}, {Wilson}, {Walmsley}  \&
  {Pauls}}{{Henkel} et~al.}{1983}]{Henkel1983}
{Henkel} C.,  {Wilson} T.~L.,  {Walmsley} C.~M.,   {Pauls} T.,  1983, \aap,
  \href {https://ui.adsabs.harvard.edu/abs/1983A&A...127..388H} {127, 388}

\bibitem[\protect\citeauthoryear{{Hern{\'a}ndez Vera}, {Lique}, {Dumouchel},
  {Hily-Blant}  \& {Faure}}{{Hern{\'a}ndez Vera}
  et~al.}{2017}]{Hernandez-Vera2017}
{Hern{\'a}ndez Vera} M.,  {Lique} F.,  {Dumouchel} F.,  {Hily-Blant} P.,
  {Faure} A.,  2017, \mn@doi [\mnras] {10.1093/mnras/stx422}, \href
  {https://ui.adsabs.harvard.edu/abs/2017MNRAS.468.1084H} {468, 1084}

\bibitem[\protect\citeauthoryear{{Iben}}{{Iben}}{1975}]{Iben1975}
{Iben} Jr. I.,  1975, \mn@doi [\apj] {10.1086/153433}, \href
  {https://ui.adsabs.harvard.edu/abs/1975ApJ...196..525I} {196, 525}

\bibitem[\protect\citeauthoryear{{Israel}, {van Dishoeck}, {Baas}, {Koornneef},
  {Black}  \& {de Graauw}}{{Israel} et~al.}{1990}]{Israel1990}
{Israel} F.~P.,  {van Dishoeck} E.~F.,  {Baas} F.,  {Koornneef} J.,  {Black}
  J.~H.,   {de Graauw} T.,  1990, \aap, \href
  {https://ui.adsabs.harvard.edu/abs/1990A&A...227..342I} {227, 342}

\bibitem[\protect\citeauthoryear{{Kameno} et~al.,}{{Kameno}
  et~al.}{2020}]{Seiji2020}
{Kameno} S.,  et~al., 2020, arXiv e-prints, \href
  {https://ui.adsabs.harvard.edu/abs/2020arXiv200409369K} {p. arXiv:2004.09369}

\bibitem[\protect\citeauthoryear{{Larson}}{{Larson}}{1981}]{Larson1981}
{Larson} R.~B.,  1981, \mn@doi [\mnras] {10.1093/mnras/194.4.809}, \href
  {https://ui.adsabs.harvard.edu/abs/1981MNRAS.194..809L} {194, 809}

\bibitem[\protect\citeauthoryear{{Liszt} \& {Lucas}}{{Liszt} \&
  {Lucas}}{2001}]{Liszt2001}
{Liszt} H.,  {Lucas} R.,  2001, \mn@doi [\aap] {10.1051/0004-6361:20010260},
  \href {https://ui.adsabs.harvard.edu/abs/2001A&A...370..576L} {370, 576}

\bibitem[\protect\citeauthoryear{{Loomis}, {Cleeves}, {{\"O}berg}, {Guzman}  \&
  {Andrews}}{{Loomis} et~al.}{2015}]{Loomis2015}
{Loomis} R.~A.,  {Cleeves} L.~I.,  {{\"O}berg} K.~I.,  {Guzman} V.~V.,
  {Andrews} S.~M.,  2015, \mn@doi [\apj] {10.1088/2041-8205/809/2/L25}, \href
  {https://ui.adsabs.harvard.edu/abs/2015ApJ...809L..25L} {809, L25}

\bibitem[\protect\citeauthoryear{{Lucas} \& {Liszt}}{{Lucas} \&
  {Liszt}}{1994}]{Lucas1994}
{Lucas} R.,  {Liszt} H.,  1994, \aap, \href
  {https://ui.adsabs.harvard.edu/abs/1994A&A...282L...5L} {282, L5}

\bibitem[\protect\citeauthoryear{{Lucas} \& {Liszt}}{{Lucas} \&
  {Liszt}}{1996}]{Lucas1996}
{Lucas} R.,  {Liszt} H.,  1996, \aap, \href
  {https://ui.adsabs.harvard.edu/abs/1996A&A...307..237L} {307, 237}

\bibitem[\protect\citeauthoryear{{Mangum} \& {Shirley}}{{Mangum} \&
  {Shirley}}{2015}]{Magnum2015}
{Mangum} J.~G.,  {Shirley} Y.~L.,  2015, \mn@doi [\pasp] {10.1086/680323},
  \href {http://adsabs.harvard.edu/abs/2015PASP..127..266M} {127, 266}

\bibitem[\protect\citeauthoryear{{McKee} \& {Ostriker}}{{McKee} \&
  {Ostriker}}{2007}]{McKee2007}
{McKee} C.~F.,  {Ostriker} E.~C.,  2007, \mn@doi [\araa]
  {10.1146/annurev.astro.45.051806.110602}, \href
  {https://ui.adsabs.harvard.edu/abs/2007ARA&A..45..565M} {45, 565}

\bibitem[\protect\citeauthoryear{{McMullin}, {Waters}, {Schiebel}, {Young}  \&
  {Golap}}{{McMullin} et~al.}{2007}]{CASA}
{McMullin} J.~P.,  {Waters} B.,  {Schiebel} D.,  {Young} W.,   {Golap} K.,
  2007, in {Shaw} R.~A.,  {Hill} F.,   {Bell} D.~J.,  eds,  Astronomical
  Society of the Pacific Conference Series Vol. 376, Astronomical Data Analysis
  Software and Systems XVI. p.~127

\bibitem[\protect\citeauthoryear{McNamara, Russell, Nulsen, Hogan, Fabian,
  Pulido  \& Edge}{McNamara et~al.}{2016}]{McNamara2016}
McNamara B.~R.,  Russell H.~R.,  Nulsen P. E.~J.,  Hogan M.~T.,  Fabian A.~C.,
  Pulido F.,   Edge A.~C.,  2016, \apj, 830, 79

\bibitem[\protect\citeauthoryear{{Meijerink}, {Spaans, M.}  \& {Israel, F.
  P.}}{{Meijerink} et~al.}{2007}]{Meijerink2007}
{Meijerink} {Spaans, M.}  {Israel, F. P.} 2007, \mn@doi [A\&A]
  {10.1051/0004-6361:20066130}, 461, 793

\bibitem[\protect\citeauthoryear{{Mittal}, {Whelan}  \& {Combes}}{{Mittal}
  et~al.}{2015}]{Mittal2015}
{Mittal} R.,  {Whelan} J.~T.,   {Combes} F.,  2015, \mn@doi [\mnras]
  {10.1093/mnras/stv754}, \href
  {http://adsabs.harvard.edu/abs/2015MNRAS.450.2564M} {450, 2564}

\bibitem[\protect\citeauthoryear{Muller, Schloder, Stutzki  \&
  Winnewisserr}{Muller et~al.}{2005}]{Splatalogue_CDMS}
Muller H. S.~P.,  Schloder F.,  Stutzki J.,   Winnewisserr G.,  2005, \mn@doi
  [Journal of Molecular Structure]
  {https://doi.org/10.1016/j.molstruc.2005.01.027}, 742, 215

\bibitem[\protect\citeauthoryear{{Muller} et~al.,}{{Muller}
  et~al.}{2011}]{Muller2011}
{Muller} S.,  et~al., 2011, \mn@doi [\aap] {10.1051/0004-6361/201117096}, \href
  {https://ui.adsabs.harvard.edu/abs/2011A&A...535A.103M} {535, A103}

\bibitem[\protect\citeauthoryear{{Muller} et~al.,}{{Muller}
  et~al.}{2013}]{Muller2013}
{Muller} S.,  et~al., 2013, \mn@doi [\aap] {10.1051/0004-6361/201220613}, \href
  {https://ui.adsabs.harvard.edu/abs/2013A&A...551A.109M} {551, A109}

\bibitem[\protect\citeauthoryear{{Nagai} et~al.,}{{Nagai}
  et~al.}{2019}]{Nagai19}
{Nagai} H.,  et~al., 2019, arXiv e-prints, \href
  {https://ui.adsabs.harvard.edu/abs/2019arXiv190506017N} {p. arXiv:1905.06017}

\bibitem[\protect\citeauthoryear{{Nawaz}, {Bicknell}, {Wagner}, {Sutherland }
  \& {McNamara}}{{Nawaz} et~al.}{2016}]{Nawaz2016}
{Nawaz} M.~A.,  {Bicknell} G.~V.,  {Wagner} A.~Y.,  {Sutherland } R.~S.,
  {McNamara} B.~R.,  2016, \mn@doi [\mnras] {10.1093/mnras/stw330}, \href
  {https://ui.adsabs.harvard.edu/abs/2016MNRAS.458..802N} {458, 802}

\bibitem[\protect\citeauthoryear{{Olivares} et~al.,}{{Olivares}
  et~al.}{2019}]{Olivares2019}
{Olivares} V.,  et~al., 2019, arXiv e-prints, \href
  {https://ui.adsabs.harvard.edu/abs/2019arXiv190209164O} {}

\bibitem[\protect\citeauthoryear{{Paglione} et~al.,}{{Paglione}
  et~al.}{2001}]{Paglione2001}
{Paglione} T. A.~D.,  et~al., 2001, \mn@doi [\apjs] {10.1086/321785}, \href
  {https://ui.adsabs.harvard.edu/abs/2001ApJS..135..183P} {135, 183}

\bibitem[\protect\citeauthoryear{{Papadopoulos}, {Seaquist}  \&
  {Scoville}}{{Papadopoulos} et~al.}{1996}]{Papadopoulos1996}
{Papadopoulos} P.~P.,  {Seaquist} E.~R.,   {Scoville} N.~Z.,  1996, \mn@doi
  [\apj] {10.1086/177410}, \href
  {https://ui.adsabs.harvard.edu/abs/1996ApJ...465..173P} {465, 173}

\bibitem[\protect\citeauthoryear{{Peng}, {Vogel}  \& {Carlstrom}}{{Peng}
  et~al.}{1995}]{Peng1995}
{Peng} Y.,  {Vogel} S.~N.,   {Carlstrom} J.~E.,  1995, \mn@doi [\apj]
  {10.1086/176570}, \href
  {https://ui.adsabs.harvard.edu/abs/1995ApJ...455..223P} {455, 223}

\bibitem[\protect\citeauthoryear{{Peterson} \& {Fabian}}{{Peterson} \&
  {Fabian}}{2006}]{PetersonFabian2006}
{Peterson} J.~R.,  {Fabian} A.~C.,  2006, \mn@doi [\physrep]
  {10.1016/j.physrep.2005.12.007}, \href
  {https://ui.adsabs.harvard.edu/abs/2006PhR...427....1P} {427, 1}

\bibitem[\protect\citeauthoryear{{Pizzolato} \& {Soker}}{{Pizzolato} \&
  {Soker}}{2005}]{Pizzolato2005}
{Pizzolato} F.,  {Soker} N.,  2005, \mn@doi [\apj] {10.1086/444344}, \href
  {http://adsabs.harvard.edu/abs/2005ApJ...632..821P} {632, 821}

\bibitem[\protect\citeauthoryear{{Qi}, {{\"O}berg}  \& {Wilner}}{{Qi}
  et~al.}{2013}]{Qi2013}
{Qi} C.,  {{\"O}berg} K.~I.,   {Wilner} D.~J.,  2013, \mn@doi [\apj]
  {10.1088/0004-637X/765/1/34}, \href
  {https://ui.adsabs.harvard.edu/abs/2013ApJ...765...34Q} {765, 34}

\bibitem[\protect\citeauthoryear{{Riquelme}, {Bronfman, L.}, {Mauersberger,
  R.}, {Finger, R.}, {Henkel, C.}, {Wilson, T. L.}  \& {Cort\'es-Zuleta,
  P.}}{{Riquelme} et~al.}{2018}]{Riquelme2017}
{Riquelme} {Bronfman, L.} {Mauersberger, R.} {Finger, R.} {Henkel, C.} {Wilson,
  T. L.}  {Cort\'es-Zuleta, P.} 2018, \mn@doi [A\&A]
  {10.1051/0004-6361/201730602}, 610, A43

\bibitem[\protect\citeauthoryear{{Roman-Duval}, {Jackson}, {Heyer}, {Rathborne}
   \& {Simon}}{{Roman-Duval} et~al.}{2010}]{RomanDuval2010}
{Roman-Duval} J.,  {Jackson} J.~M.,  {Heyer} M.,  {Rathborne} J.,   {Simon} R.,
   2010, \mn@doi [\apj] {10.1088/0004-637X/723/1/492}, \href
  {https://ui.adsabs.harvard.edu/abs/2010ApJ...723..492R} {723, 492}

\bibitem[\protect\citeauthoryear{Rose et~al.,}{Rose et~al.}{2019a}]{Rose2019a}
Rose T.,  et~al., 2019a, \mn@doi [\mnras] {10.1093/mnras/stz406}, 485, 229

\bibitem[\protect\citeauthoryear{Rose et~al.,}{Rose et~al.}{2019b}]{Rose2019b}
Rose T.,  et~al., 2019b, \mn@doi [\mnras] {10.1093/mnras/stz2138}, 489, 349

\bibitem[\protect\citeauthoryear{Ruffa et~al.,}{Ruffa et~al.}{2019}]{Ruffa2019}
Ruffa I.,  et~al., 2019, \mn@doi [\mnras] {10.1093/mnras/stz255}, 484, 4239

\bibitem[\protect\citeauthoryear{Russell, McNamara, Edge, Hogan, Main  \&
  Vantyghem}{Russell et~al.}{2013}]{Russell2013}
Russell H.~R.,  McNamara B.~R.,  Edge A.~C.,  Hogan M.~T.,  Main R.~A.,
  Vantyghem A.~N.,  2013, \mn@doi [\mnras] {10.1093/mnras/stt490}, 432, 530

\bibitem[\protect\citeauthoryear{Sliwa, Wilson, Matsushita, Peck, Petitpas,
  Saito  \& Yun}{Sliwa et~al.}{2017}]{Sliwa2017}
Sliwa K.,  Wilson C.~D.,  Matsushita S.,  Peck A.~B.,  Petitpas G.~R.,  Saito
  T.,   Yun M.,  2017, \mn@doi [\apj] {10.3847/1538-4357/aa689b}, 840, 8

\bibitem[\protect\citeauthoryear{{Solomon}, {Rivolo}, {Barrett}  \&
  {Yahil}}{{Solomon} et~al.}{1987}]{Solomon1987}
{Solomon} P.~M.,  {Rivolo} A.~R.,  {Barrett} J.,   {Yahil} A.,  1987, \mn@doi
  [\apj] {10.1086/165493}, \href
  {https://ui.adsabs.harvard.edu/abs/1987ApJ...319..730S} {319, 730}

\bibitem[\protect\citeauthoryear{Taniguchi, Ohyama  \& Sanders}{Taniguchi
  et~al.}{1999}]{Taniguchi1999}
Taniguchi Y.,  Ohyama Y.,   Sanders D.~B.,  1999, \mn@doi [\apj]
  {10.1086/307650}, 522, 214

\bibitem[\protect\citeauthoryear{{Taylor}}{{Taylor}}{1996}]{Taylor1996}
{Taylor} G.~B.,  1996, \mn@doi [\apj] {10.1086/177874}, \href
  {http://adsabs.harvard.edu/abs/1996ApJ...470..394T} {470, 394}

\bibitem[\protect\citeauthoryear{{Taylor}, {Perley}, {Inoue}, {Kato}, {Tabara}
  \& {Aizu}}{{Taylor} et~al.}{1990}]{Taylor1990}
{Taylor} G.~B.,  {Perley} R.~A.,  {Inoue} M.,  {Kato} T.,  {Tabara} H.,
  {Aizu} K.,  1990, \mn@doi [\apj] {10.1086/169094}, \href
  {http://adsabs.harvard.edu/abs/1990ApJ...360...41T} {360, 41}

\bibitem[\protect\citeauthoryear{Temi, Amblard, Gitti, Brighenti, Gaspari,
  Mathews  \& David}{Temi et~al.}{2018}]{Temi2018}
Temi P.,  Amblard A.,  Gitti M.,  Brighenti F.,  Gaspari M.,  Mathews W.~G.,
  David L.,  2018, \apj, 858, 17

\bibitem[\protect\citeauthoryear{Tremblay et~al.}{Tremblay
  et~al.}{2016}]{Tremblay2016}
Tremblay G.~R.,  et~al., 2016, \mn@doi [Nature] {10.1038/nature17969}, 534, 218

\bibitem[\protect\citeauthoryear{{Tremblay} et~al.,}{{Tremblay}
  et~al.}{2018}]{Tremblay2018}
{Tremblay} G.~R.,  et~al., 2018, \mn@doi [\apj] {10.3847/1538-4357/aad6dd},
  \href {http://adsabs.harvard.edu/abs/2018ApJ...865...13T} {865, 13}

\bibitem[\protect\citeauthoryear{{Vantyghem} et~al.,}{{Vantyghem}
  et~al.}{2017}]{Vantyghem2017}
{Vantyghem} A.~N.,  et~al., 2017, \mn@doi [\apj] {10.3847/1538-4357/aa8fd0},
  \href {https://ui.adsabs.harvard.edu/abs/2017ApJ...848..101V} {848, 101}

\bibitem[\protect\citeauthoryear{{Vazquez-Semadeni}, {Gomez}, {Jappsen},
  {Ballesteros-Paredes}, {Gonzalez}  \& {Klessen}}{{Vazquez-Semadeni}
  et~al.}{2007}]{Vazquez-Semadeni2007}
{Vazquez-Semadeni} E.,  {Gomez} G.~C.,  {Jappsen} A.~K.,  {Ballesteros-Paredes}
  J.,  {Gonzalez} R.~F.,   {Klessen} R.~S.,  2007, \mn@doi [\apj]
  {10.1086/510771}, \href
  {https://ui.adsabs.harvard.edu/abs/2007ApJ...657..870V} {657, 870}

\bibitem[\protect\citeauthoryear{{Voit}, {Donahue}, {Bryan}  \&
  {McDonald}}{{Voit} et~al.}{2015}]{Voit2015}
{Voit} G.~M.,  {Donahue} M.,  {Bryan} G.~L.,   {McDonald} M.,  2015, \mn@doi
  [\nat] {10.1038/nature14167}, \href
  {https://ui.adsabs.harvard.edu/abs/2015Natur.519..203V} {519, 203}

\bibitem[\protect\citeauthoryear{{Wiklind} \& {Combes}}{{Wiklind} \&
  {Combes}}{1996a}]{Wiklind1996b}
{Wiklind} T.,  {Combes} F.,  1996a, \aap, \href
  {https://ui.adsabs.harvard.edu/abs/1996A&A...315...86W} {315, 86}

\bibitem[\protect\citeauthoryear{{Wiklind} \& {Combes}}{{Wiklind} \&
  {Combes}}{1996b}]{Wiklind1996a}
{Wiklind} T.,  {Combes} F.,  1996b, \mn@doi [\nat] {10.1038/379139a0}, \href
  {https://ui.adsabs.harvard.edu/abs/1996Natur.379..139W} {379, 139}

\bibitem[\protect\citeauthoryear{{Wiklind} \& {Combes}}{{Wiklind} \&
  {Combes}}{1997a}]{Wiklind1997a}
{Wiklind} T.,  {Combes} F.,  1997a, \aap, \href
  {https://ui.adsabs.harvard.edu/abs/1997A&A...324...51W} {324, 51}

\bibitem[\protect\citeauthoryear{{Wiklind} \& {Combes}}{{Wiklind} \&
  {Combes}}{1997b}]{Wiklind1997b}
{Wiklind} T.,  {Combes} F.,  1997b, \aap, \href
  {https://ui.adsabs.harvard.edu/abs/1997A&A...328...48W} {328, 48}

\bibitem[\protect\citeauthoryear{{Wilson}}{{Wilson}}{1999}]{Wilson1999}
{Wilson} T.~L.,  1999, \mn@doi [Reports on Progress in Physics]
  {10.1088/0034-4885/62/2/002}, \href
  {https://ui.adsabs.harvard.edu/abs/1999RPPh...62..143W} {62, 143}

\bibitem[\protect\citeauthoryear{Wilson}{Wilson}{2018}]{Wilson2018}
Wilson C.~D.,  2018, \mn@doi [\mnras] {10.1093/mnras/sty845}, 477, 2926

\bibitem[\protect\citeauthoryear{{Zhang}, {Romano}, {Ivison}, {Papadopoulos}
  \& {Matteucci}}{{Zhang} et~al.}{2018}]{Zhang2018}
{Zhang} Z.-Y.,  {Romano} D.,  {Ivison} R.~J.,  {Papadopoulos} P.~P.,
  {Matteucci} F.,  2018, \mn@doi [\nat] {10.1038/s41586-018-0196-x}, \href
  {https://ui.adsabs.harvard.edu/abs/2018Natur.558..260Z} {558, 260}

\bibitem[\protect\citeauthoryear{{van de Voort}, {Schaye}, {Altay}  \&
  {Theuns}}{{van de Voort} et~al.}{2012}]{voort2012}
{van de Voort} F.,  {Schaye} J.,  {Altay} G.,   {Theuns} T.,  2012, \mn@doi
  [\mnras] {10.1111/j.1365-2966.2012.20487.x}, \href
  {https://ui.adsabs.harvard.edu/abs/2012MNRAS.421.2809V} {421, 2809}

\makeatother
\end{thebibliography}
\end{document}